\documentclass[twocolumn,prl,amsmath,amssymb,10pt,aps]{revtex4-1} 
\usepackage{graphicx}
\usepackage{latexsym}
\usepackage{amsmath}
\usepackage{amsthm}
\usepackage{amssymb}
\usepackage{epstopdf} 
\usepackage{enumerate}
\usepackage{setspace} 
\usepackage{dcolumn}
\usepackage{bm}
\usepackage{setspace} 
\usepackage{slashed}
\usepackage{color}
\usepackage{youngtab}
\usepackage[colorlinks=true]{hyperref}
\usepackage{float}
\usepackage{graphicx}
\usepackage{multirow}
\usepackage{array}
\newcolumntype{L}[1]{>{\raggedright\arraybackslash}p{#1}}
\newcolumntype{C}[1]{>{\centering\arraybackslash}p{#1}}
\newcolumntype{R}[1]{>{\raggedleft\arraybackslash}p{#1}}
\usepackage{tikz}
\usepackage{tikz-3dplot}
\usepackage{braket}
\usepackage{cancel}
\usepackage{makecell}
\usepackage{subfigure}
\usetikzlibrary{decorations.pathreplacing,calc}
\usetikzlibrary{shapes,snakes,arrows,chains,matrix,positioning,scopes,calc}
\usetikzlibrary{decorations.markings}
\tikzset{/pgf/decoration/.cd,
    number of sines/.initial=10,
    angle step/.initial=20,
}
\newif\ifmirrorsemicircle

\tikzset{
    wave amplitude/.initial=0.2cm,
    wave count/.initial=8,
    mirror semicircle/.is if=mirrorsemicircle,
    mirror semicircle=false,
    wavy semicircle/.style={
        to path={
            let \p1 = (\tikztostart),
            \p2 = (\tikztotarget),
            \n1 = {veclen(\y2-\y1,\x2-\x1)},
            \n2 = {atan2(\y2-\y1,\x2-\x1))} in
            plot [
                smooth,
                samples=(\pgfkeysvalueof{/tikz/wave count}+0.5)*8+1, 
                domain=0:1,
                shift={($(\p1)!0.5!(\p2)$)}
            ] ({ 
                (\x*180-\n2 + 180 + \ifmirrorsemicircle 1 \else -1 \fi * 90%
            }:{ 
                (%
                    \n1/2+\pgfkeysvalueof{/tikz/wave amplitude} * %
                    sin(
                        \x * 360 * (\pgfkeysvalueof{/tikz/wave count} + 0.5%
                    )%
                )%
            })
        } (\tikztotarget)
    }
}

\newdimen\tmpdimen
\pgfdeclaredecoration{complete sines}{initial}
{
    \state{initial}[
        width=+0pt,
        next state=move,
        persistent precomputation={
            \pgfmathparse{\pgfkeysvalueof{/pgf/decoration/angle step}}
            \let\anglestep=\pgfmathresult%
            \let\currentangle=\pgfmathresult%
            \pgfmathsetlengthmacro{\pointsperanglestep}%
                {(\pgfdecoratedremainingdistance/\pgfkeysvalueof{/pgf/decoration/number of sines})/360*\anglestep}%
        }] {}
    \state{move}[width=+\pointsperanglestep, next state=draw]{
        \pgfpathmoveto{\pgfpointorigin}
    }
    \state{draw}[width=+\pointsperanglestep, switch if less than=1.25*\pointsperanglestep to final, 
        persistent postcomputation={
        \pgfmathparse{mod(\currentangle+\anglestep, 360)}%
        \let\currentangle=\pgfmathresult%
    }]{%
        \pgfmathsin{+\currentangle}%
        \tmpdimen=\pgfdecorationsegmentamplitude%
        \tmpdimen=\pgfmathresult\tmpdimen%
        \divide\tmpdimen by2\relax%
        \pgfpathlineto{\pgfqpoint{0pt}{\tmpdimen}}%
    }
    \state{final}{
        \ifdim\pgfdecoratedremainingdistance>0pt\relax
            \pgfpathlineto{\pgfpointdecoratedpathlast}
        \fi
   }
}
\allowdisplaybreaks 
\usepackage{color}

\begin{document}
\title{ Multiscale Quantum Criticality Driven by\\
Kondo Lattice Coupling in Pyrochlore Systems  }

\author{Hanbit Oh$^1$}
\author{Sangjin Lee$^1$}
\author{Yong Baek Kim$^2$}
\thanks{ybkim@physics.toronto.ca}
\author{Eun-Gook Moon$^1$}
\thanks{egmoon@kaist.ac.kr}

\affiliation{$^1$Department of Physics, Korea Advanced Institute of Science and Technology, Daejeon 305-701, Korea}
\affiliation{$^2$Department of Physics and Centre for Quantum Materials, University of Toronto, Toronto, Ontario M5S 1A7, Canada}
\date{\today}
\textcolor{red}{}

\begin{abstract}   
Pyrochlore systems ($A_{2}B_{2}O_{7}$) with $A$-site rare-earth local moments and $B$-site $5d$ conduction electrons offer excellent material platforms for the discovery of exotic quantum many-body ground states. Notable examples include U(1) quantum spin liquid of the local moments and semimetallic non-Fermi liquid of the conduction electrons. 
Here we investigate emergent quantum phases and their transitions driven by the Kondo lattice coupling between such highly entangled quantum ground states. 
Using the renormalization group method, it is shown that weak Kondo lattice coupling is irrelevant, leading to a fractionalized semimetal phase with decoupled local moments and conduction electrons. 
Upon increasing the Kondo lattice coupling, this phase is unstable to the formation of broken symmetry states. Particularly important is the opposing influence of the Kondo lattice coupling and long-range Coulomb interaction. The former prefers to break the particle-hole symmetry while the latter tends to restore it. The characteristic competition leads to possibly multiple phase transitions, first from a fractionalized semimetal phase to a fractionalized Fermi surface state with particle-hole pockets, followed by the second transition to a fractionalized ferromagnetic state. Multiscale quantum critical behaviors appear at nonzero temperatures and with   external magnetic field near such quantum phase transitions. We discuss the implication of these results to the experiments on Pr$_2$Ir$_2$O$_7$. 
 \end{abstract}

\maketitle

 \textit{Introduction} : 
 Recent advances in correlated electron systems reveal emergent phenomena beyond the Landau paradigm. Localized magnetic moments may host quantum spin liquid (QSL) phases characterized by fluctuating gauge fields and fractionalized particles \cite{Gingras, Balents, Zhou}. Itinerant electron systems may show non-Fermi liquid behavior without quasiparticles \cite{NFL, LeeSS}. Such phenomena and associated quantum phase transitions demand development of new concepts and novel understandings in strongly interacting quantum many-body systems \cite{deconfined,  Wen,   YB}.

In this work, we study the intertwined model of two emergent phases beyond the Landau paradigm. We consider the interaction between the local moment system supporting a U(1) quantum spin liquid and a non-Fermi liquid semimetallic state of conduction electrons. This model is partly motivated by physics of the pyrochlore materials,  $A_{2}B_{2}O_{7}$, where the $A$- and $B$-site pyrochlore lattices are occupied by rare-earth local moments and $5d$ conduction electrons, respectively. The $A$-site local moments may form a quantum spin liquid with  emergent photons, also known as quantum spin-ice, as suggested in Yb$_2$Ti$_2$O$_7$, Pr$_2$Hf$_2$O$_7$, and Pr$_2$Zr$_2$O$_7$ \cite{Kimura, Ong, Ross, Tokiwa1, Armitage, Petrenko, Lake, Coldea, Tokiwa2}. When the $B$-site is occupied by conduction electrons in $5d$ orbitals, such as J$_{\rm eff}$=1/2 Kramers doublet of Ir ions, the system supports a non-Fermi liquid semimetal, the so-called Luttinger-Abrikosov-Beneslaevski (LAB) state, which is derived from the quadratic band touching with long-range Coulomb interaction \cite{Luttinger, Abrikosov1, Abrikosov2, Moon}. The quadratic band touching of Ir conduction electrons is confirmed in the angle-resolved photoemission spectroscopy (ARPES) experiment in the finite-temperature paramagnetic state of Pr$_2$Ir$_2$O$_7$ and Nd$_2$Ir$_2$O$_7$ \cite{Kondo1,Kondo2}. It is believed that the interplay between the two emergent phases mentioned above may play a crucial role in low-temperature physics of Pr$_2$Ir$_2$O$_7$ \cite{Armitage2, Nakatsuji1, Nakatsuji2, Tokiwa3} .

\begin{figure}
\centering
	\includegraphics[width=\linewidth]{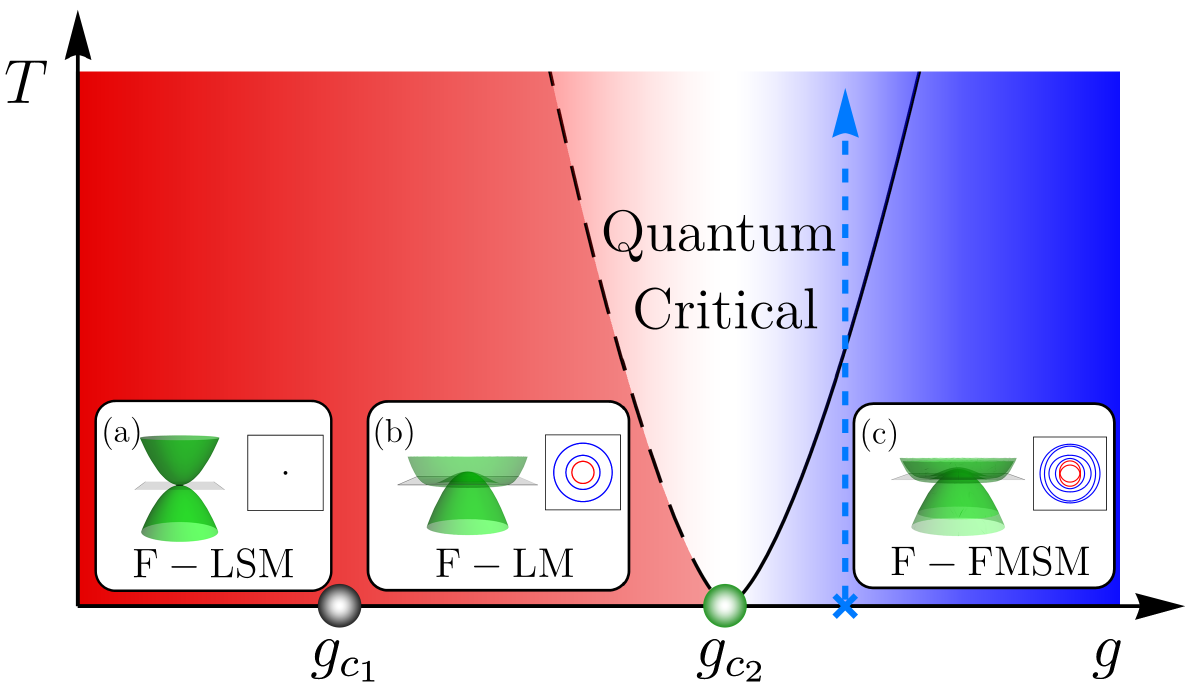}
	\caption{Schematic phase diagram. The coupling constant $g$ characterizes strength of the Kondo lattice coupling. The particle-hole symmetry is broken at $g_{c1}$ and the time reversal is broken at $g_{c2}$. Insets show energy dispersions of the conducting electrons.  Systems at the blue dashed line may show multicritical behaviors with a nonzero magnetic moment at low temperatures.}
	\label{Phase}
\end{figure}

Considering the Kondo lattice coupling between the localized moments and conduction electrons, we first construct a low energy effective field theory for the coupled system of the U(1) QSL and the non-Fermi liquid semimetal state. 
We use the renormalization group method to investigate emergent phases and phase transitions and find that the Kondo lattice coupling and the long-range Coulomb interaction show intriguing interplay physics. Namely, the former shows the tendency of breaking the particle-hole symmetry but the latter plays the opposite role. 
For small Kondo lattice coupling, the Coulomb interaction prevails, and the two underlying phases remain weakly coupled. This is the Luttinger semimetal  coexisting with fractionalized excitations and emergent photons from the local moments. 
For sufficiently strong interaction, either time reversal or inversion symmetry may be broken and our perturbative renormalization group analysis shows that time reversal symmetry breaking is the most relevant channel. Our analysis suggests that the particle-hole symmetry breaking occurs first as the Kondo lattice coupling becomes dominant over the long-range Coulomb interaction, leading to a fractionalized Fermi surface state with emergent particle-hole pockets. This is followed by the time reversal symmetry breaking transition to a ferromagnetically ordered fractionalized semimetal phase. We discuss the resulting multiscaling critical behavior in light of some key experiments in the low-temperature phase of Pr$_2$Ir$_2$O$_7$.

\textit{Model} :
We start with a generic model Hamiltonian for the pyrochlore system,  $A_{2}B_{2}O_{7}$,  
\begin{eqnarray}
H_{tot} &=& H_A + H_B + H_{A-B} \nonumber \\
H_A &=& \sum_{\langle v,u \rangle} J_{\mu \nu} (u, v) S^{\mu}(u) S^{\nu}(v)  \nonumber \\
H_B &=&   - \sum_{\langle i,j \rangle} t_{ij}^{\alpha \beta} f_{\alpha}^{\dagger} (i)  f_{\beta}(j) + \frac{e^2}{2} \sum_{i \neq j}\frac{n_B(i) n_B(j)}{|i-j|} \nonumber \\
H_{A-B}&=& \sum_{i,t} R_{\mu \nu}(v,i) S^{\mu}(v)  \big( f_{\alpha}^{\dagger} (i) \,s_{\alpha \beta}^{\nu} \, f_{\beta}(i) \big). \nonumber
\end{eqnarray}
The Hamiltonian for $A$ sites ($H_A$) describes localized magnetic moments ($S^{\mu}$), and the Hamiltonian for $B$ sites ($H_B$) describes conduction electrons with annihilation and creation operators ($f_{\alpha}, f_{\beta}^{\dagger}$). Greek indices ($\alpha, \beta$) are for spin quantum numbers, and ($u, v$) and ($i,j$) are indices for $A$ and $B$ sites, respectively. 
The Kondo lattice coupling is described by $H_{A-B}$.
A generic hopping term ($t_{ij}^{\alpha \beta}$),  generic exchange interaction $\left[J_{\mu \nu} (u,v)\right]$, and interaction function $\left[R_{\mu \nu}(u,i), s_{\alpha \beta}^{\nu}\right]$ are introduced, which are constrained by lattice symmetry. The long-range Coulomb interaction with electric charge $e$ is also introduced.
The local spin operators may be represented in terms of either global or local axes, 
$\vec{S}(u) = S^{a}(u) \hat{e}_{a}(u) =S^{\mu}(u) \hat{x}_{\mu}$. The basis vectors of the local axes ($ \hat{e}_{a}(u)$) are commonly used in spin-ice literature \cite{KimYB, Hermele, Shannon, Savary2, SBLee}, and it is straightforward to find the relations with the basis vectors of the global axes ($ \hat{x}_{\mu}$).

 We focus on the system where spins at $A$ sites host a U(1) QSL and electrons at $B$ sites form the Luttinger semi-metal, motivated by the ARPES experiments \cite{Kondo1}.  
 The low energy effective Hamiltonian of $H_A$ may be written as the quantum spin-ice Hamiltonian, $ H_A \rightarrow \sum_{v} \frac{1}{2\mu_0} \vec{\mathcal{B}}(v))^2 + \sum_{v^*} \frac{\epsilon_0}{2} (\vec{\mathcal{E}}(v^*))^2$ \cite{KimYB, Hermele, Shannon}. 
The ``star" index, $v^*$, represents the dual diamond lattice sites of the underlying $A$-site pyrochlore lattice. 
The emergent magnetic field $\vec{\mathcal{B}}(v)$ is proportional to the average of the local spin projections to easy-axis ([111] or equivalent) directions, 
and the emergent electric field ($\vec{\mathcal{E}}(v^*)$) describes spin fluctuations out of local easy-axis directions. We stress that the quantum spin-ice manifold is defined by the divergence-free condition, $\nabla \cdot \vec{\mathcal{B}}=0$, and thus the $\vec{\mathcal{B}}$ fluctuations are all transversal. 

The Luttinger semi-metal Hamiltonian  can be approximated as $H_B(e=0)  \rightarrow  \sum_{\vec{k}} \Psi^{\dagger}_{k} \mathcal{H}_0(\vec{k}) \Psi_k$ with a four component spinor $\Psi_{k}$, 
 \begin{eqnarray}
  \mathcal{H}_{0}(\vec{k}) &=& \frac{c_{0}}{2m}\vec{k}^{2}+\frac{c_{1}}{2m}\sum_{n=1}^{3}d_{n}(\vec{k})\Gamma^{n}+\frac{c_{2}}{2m}\sum_{n=4}^{5}d_{n}(\vec{k})\Gamma^{n}. \nonumber
 \end{eqnarray} 
 At low energy, the functions $d_n({\vec k})$ may be written as  
 \begin{eqnarray}
d_{1}(\vec{k})=\sqrt{3}k_{x}k_{y},\  d_{2}(\vec{k})=\sqrt{3}k_{x}k_{z},\ d_{3}(\vec{k})=\sqrt{3}k_{y}k_{z}\nonumber\\
d_{4}(\vec{k})=\frac{\sqrt{3}}{2}\left(k_{x}^2-k_{y}^{2}\right),\  d_{5}(\vec{k})=\frac{1}{2 }\left(2k_{z}^2-k_{x}^2-k_{y}^{2}\right). \nonumber
\end{eqnarray}
The term with $c_{0}$ breaks the particle-hole symmetry (PHS), and the terms with  $c_{1},c_{2}$ are associated with the $t_{2g}$ and $e_{g}$ representations, respectively, of the cubic symmetry. 
We emphasize that the charge neutrality of the system would demand the particle-hole band condition  $|c_{0}| < |c_{1}|, |c_{2}|$ if there are no other electron-hole pockets far away from the zone center\cite{Kondo1}. We also set $m=1/2$ for simplicity unless otherwise stated.   
Note that the presence of the long-range Coulomb interaction makes the particle-hole and SO(3) rotational symmetries emergent in the LAB phase ($c_0 \rightarrow 0$ and $c_1 \rightarrow c_2$)\cite{Moon}.

The two sectors ($A,B$) host the excitations with fundamentally different dynamics. Namely, low energy excitations of the $A$-sites are emergent photons whose dispersion relation is $\omega_{\lambda}(\vec{q}) = v_{\lambda} |\vec{q}|$ with two polarizations $\lambda$ and their velocities $v_{\lambda}$. 
On the other hand,  the $B$-sites have electronic excitations with the long-range Coulomb interaction $\epsilon(\vec{k}) \sim |\vec{k}|^z$ with $z \sim 2$. 

We first construct a low energy effective coupling term of the Kondo lattice coupling. Employing the lattice symmetries and gauge invariance, the lowest order coupling terms may be written as
\begin{eqnarray}
H_{A-B} \rightarrow   g \int_x   \mathcal{B}_i  \Psi^{\dagger} \hat{ {M}}_i \Psi,\quad \hat{{M}}_i = \cos(\alpha) s_i + \sin(\alpha) (s_i)^3. \nonumber
 \end{eqnarray}
The two coupling constants ($g$ and $\alpha$) characterize the Kondo lattice coupling. 
The specific form of the coupling matrix $\hat{ {M}} $ is completely determined by the cubic and time reversal symmetries.
The 4$\times$4 matrix ($ s_i $) of a spin operator $S_i \equiv \Psi^{\dagger} s_i \Psi$ is used with the explicit form being introduced in Supplemental Material(SM). 
We mainly focus on the coupling to $\vec{\mathcal{B}}$ because $\vec{\mathcal{E}}$ couples to conduction electrons through a polarization-type coupling $\hat{{M}}_{E}=\nabla$ or a Rashba-type coupling $\hat{ {M}}_{E}=\nabla \times \vec{s}$, which is less relevant than the coupling to $\vec{\mathcal{B}}$ (see SM). 
 
 The effective low energy action of the total Hamiltonian in the Euclidean spacetime is 
\begin{eqnarray}
&&\mathcal{S}_{tot} =\mathcal{S}_{A}+\mathcal{S}_{B}+\mathcal{S}_{A-B}\nonumber \\
 &&\mathcal{S}_{A} = \int_{x, \tau}
 \frac{(\mathcal{\vec{B}})^2 }{2\mu_0} - \frac{\epsilon_0 (\mathcal{\vec{E}})^2 }{2}   + O((\nabla_i \mathcal{B}_j )^2, (\nabla_i \mathcal{E}_j )^2, \mathcal{E}_i^4, \mathcal{B}_j^4 ) \nonumber\\
&&\mathcal{S}_{B} = \int_{x,\tau} \Psi^{\dagger}\left(\partial_{\tau}+\mathcal{H}_{0}(-i \nabla)\right)\Psi +\frac{e^2}{2} \int_{x,y,\tau}  \frac{n(x,\tau) n(y,\tau)}{|x-y|}\nonumber \\
&& \mathcal{S}_{A-B} = g \int_{x,\tau} \,  \mathcal{B}_i \Psi^{\dagger} \hat{ {M}}_i \Psi, 
\nonumber
\end{eqnarray}
with the density operator, $n(x,\tau)= \Psi^{\dagger}(x,\tau) \Psi(x,\tau)$.
Though  the form of the Yukawa coupling with $g$ may look similar to the ones in previous literatures \cite{Savary, Murray, ChenG}, we emphasize that the U(1) gauge structure in the spin-ice manifold plays a crucially different role here.

With a nonzero small coupling ($g \neq 0$), the dynamics of the $\mathcal{B}$ fields is modified as 
\begin{eqnarray}
\int_q \Big(\frac{\delta_{ij}}{\mu_0} \Big)\frac{\mathcal{B}_i (-q)  \mathcal{B}_j (q)}{2} \,\, \rightarrow \,\, \int_q  \Big(\frac{\delta^{ij}}{\mu_0} + \Sigma_{\mathcal{B}}^{ij}(q) \Big) \frac{\mathcal{B}_i (-q)  \mathcal{B}_j (q)}{2}. \nonumber
\end{eqnarray}
Hereafter, we use the four-vector notation for momentum and frequency, $k \equiv (\vec{k}, k_n)$. 
The boson self-energy at one-loop order is   
\begin{eqnarray}
\Sigma_{\mathcal{B}}^{ij}(q) =g^2 \int_{\vec{k},k_{n}}{\rm Tr}\big(\hat{\mathcal{M}}_{i} {G}^{0}_f (k+q)\hat{\mathcal{M}}_{j}  {G}^{0}_f(k)\big) 
 \end{eqnarray} 
with the Fermion Green's function $G^{0}_f(k) =( -i k_n + \mathcal{H}_0(\vec{k}))^{-1}$.
 Introducing an ultraviolet (UV) cutoff $\Lambda$, we find 
$\Sigma_{\mathcal{B}}^{ij}(q=0) = -  \delta^{ij}\frac{g^2\Lambda }{2\pi^2}$
for the most symmetric condition ($c_1=c_2=1$, $c_0=0$, and $\alpha=0$), which corresponds to  $\frac{1}{\mu_0} \rightarrow \frac{1}{\mu_0} (1- \frac{\mu_0 g^2\Lambda }{2\pi^2} )  $.
The fermion self-energy from the gauge fluctuations is 
\begin{eqnarray}
\Sigma_{f}(k)&=&-{g}^{2}\int_{q} \hat{M}_{i} \, G_{f}^0(k+q) \, \hat{M}_{j} \, \langle \mathcal{B}_i (-q) \mathcal{B}_j(+q) \rangle. 
\end{eqnarray}
Unless $1/\mu_0 =0$, we find the fermion self-energy,
\begin{eqnarray}
\Sigma_f(k) = \mu_0 g^2 \Lambda  \Big( \delta c_0 \frac{\vec{k}^2}{2m}   +  \frac{\delta c_{1}}{2m}\sum_{n=1}^{5}d_{n}(\vec{k})\Gamma^{n} \Big) +\cdots \label{self}
\end{eqnarray}
with nonzero values of $\delta c_0, \delta c_1$. For the most symmetric condition, we find $\delta c_0=0.025$, $\delta c_1=-0.013$.
The two self-energies are UV divergent under the Kondo lattice coupling. 
The absence of logarithmic dependence on the UV cutoff means the bosonic and fermionic excitations remain weakly coupled, and the decoupled ground state ($g=0$) is stable as far as the Kondo lattice coupling is small. We call this state the fractionalized Luttinger semimetal (F-LSM). 

We stress that the Kondo lattice coupling generates the  PHS breaking term as manifested by $\delta c_0 \neq 0$ even with   the most symmetric bare Hamiltonian ($c_0=0$). 
This contribution arises from the transversal gauge fluctuations, and the absence of the longitudinal gauge fluctuations is essential. 
The competition between the Kondo lattice coupling and the long-range Coulomb interaction plays an important role in the PHS channel of F-LSM.  
 In the perturbative regime $\mu_0 g^2 \Lambda \ll 1$, the long-range Coulomb interaction is more relevant than the Kondo lattice coupling \cite{Moon}, and the PHS is realized.

Adjacent phases of F-LSM may be obtained by using the lattice symmetries.  
Considering time reversal symmetry and parity as well as the rotational symmetries, a ground state with $\langle \mathcal{B}_i \rangle \neq 0$ breaks time reversal symmetry  and rotations but not parity. The resulting time reversal broken phase hosts nodal conduction electrons in the form of Weyl semi-metals or metals. We call such a semi-metal phase with broken time reversal symmetry a fractionalized ferromagnetic semi-metal (F-FMSM). The presence of both gapless electronic and gauge excitations is one of the main characteristics of F-FMSM, different from other states such as the Coulombic ferromagnetic state \cite{Savary2}.
Similarly, a state with $\langle \mathcal{E}_i \rangle \neq 0$ and $\langle \mathcal{B}_i \rangle = 0$ is naturally dubbed  a fractionalized ferroelectric phase, and one with $\langle \mathcal{E}_i \rangle, \langle \mathcal{B}_i \rangle \neq 0$ may be called a fractionalized multiferroic phase. As shown below, however, 
there may exist another transition before the system reaches the F-FMSM.

\textit{Quantum phase transitions} :
To investigate transitions to symmetry-broken phases,  we first extend and apply Landau's mean- field analysis, which amounts to ignoring spatial and temporal fluctuations of the emergent fields via $ \mathcal{B}_i(x,\tau) \rightarrow \mathcal{B}_i$. Integrating out the fermion excitations, we obtain the effective action of $\mathcal{B}_i$, and a continuous quantum phase transition between F--LSM and F-FMSM is obtained for $g > g_{c}$ (similar to the one of Ref. \cite{Murray}, and also see SM).
At the one-loop level, we find $  g_{c}= \sqrt{  2\pi^2 /( \mu_0 \Lambda)}$ for the most symmetric condition. 
The continuous transition obtained in the mean-field calculation respects the PHS and SO(3) rotational symmetry because F-LSM enjoys those symmetries.  

We, however, show that the gauge fluctuations destabilize the continuous transition of the mean-field calculation.  
Defining $\delta \Sigma_{\mathcal{B}}^{ij}(q) = \Sigma_{\mathcal{B}}^{ij} (q) - \Sigma_{\mathcal{B}}^{ij}(0)$, we find that the boson self-energy has the form,
\begin{eqnarray}
\delta \Sigma_{\mathcal{B}}^{ij}(q) =  g^2 \Big( a_T |\vec{q}|+a_{\omega} \sqrt{|q_n|} \Big)(\delta^{ij}- \frac{q^i q^j}{\vec{q}^2})  \nonumber. \label{bself}
\end{eqnarray}
 The condition $\nabla \cdot \mathcal{B}=0$ enforces that the fluctuations are transversal, and the dimensionless functions ($a_T, a_{\omega}$) are positive in a wide range of the parameters which are illustrated in SM. 
 Their positiveness indicates that the gauge fluctuations are stable with the renormalized propagator $ \big(\Sigma_{\mathcal{B}}^{ij} (q)\big)^{-1} $. 

Let us assume that there is a stable continuous transition between F-LSM and F-FMSM. 
At the critical point ($g=g_c$), the dominant boson propagator may be written as 
\begin{eqnarray}
&&\langle \mathcal{B}_i (-q) \mathcal{B}_j(+q) \rangle   =\frac{1}{g_{c}^2} \frac{1}{  a^T |\vec{q}| +a_{\omega} \sqrt{|q_n|} }(\delta^{ij}- \hat{q}^i \hat{q}^j) \label{boson} 
\end{eqnarray}
omitting higher order terms. To control calculations better, one may introduce the flavor number of fermions ($N_f$) and perform $1/N_f$ calculations (see SM).
The PHS breaking term can be obtained by evaluating 
\begin{eqnarray}
\frac{\partial {\rm Tr} \big[\Sigma_{f}(k)\big]}{\partial \vec{k}^2}&=&- \int_{q}  \, \frac{\partial }{\partial \vec{k}^2}  \, \frac{{\rm Tr} \big[G_{f}^0(k+q)(\hat{M}_{i} \hat{M}_{i}-\hat{M}_{i}  \hat{M}_{j}  \hat{q}^i \hat{q}^j ) \big]}{  a^T |\vec{q}| +a_{\omega} \sqrt{|q_n|} } . \nonumber 
\end{eqnarray}
The integral is logarithmically divergent,
\begin{eqnarray}
\Lambda \frac{\partial}{\partial \Lambda} \Big( \frac{\partial  {\rm Tr}\big[ \Sigma_f(k) \big]}{\partial \vec{k}^2} \Big)=4 \delta_0.  \label{PH}
\end{eqnarray}
Including both the gauge fluctuations and the long-range Coulomb interaction, we find $\delta_0 \neq 0$ ($\delta_0=0.3601$ for the most symmetric condition), which can be interpreted as a divergent $\delta c_0$.
The logarithmic divergence demonstrates the PHS cannot be realized at the critical point, which indicates that the Kondo lattice coupling dominates the long-range Coulomb interaction near the critical point. Thus, there is no continuous quantum phase transition between F-LSM and F-FMSM.

The divergence of the PHS breaking term destabilizes not only the validity of the mean-field calculation but also the particle-hole band condition ($|c_0| < |c_{1}|, |c_2|$). We find that the corrections of $c_{1,2}$ are smaller than the one of $c_0$, and thus  the particle-hole symmetry condition may break down at long wavelength and low energy. The charge neutrality condition then enforces the formation of electron and hole pockets near the Brillouin zone center.

We propose, based on the above calculations, that the PHS is broken before the onset of $\langle \mathcal{B}_i \rangle$, whose validity is  self-consistently checked \textit{a posteriori}. The transition between F-LSM and F-FMSM is intervened by an intermediate phase with the electron and hole pockets dubbed the fractionalized Luttinger metal (F-LM).  
There {\it must} be more than one continuous transition between F-LSM and F-FMSM as illustrated in Fig. 1. The transition between F-LSM and F-LM is likely to be described by the Lifshitz transition. Once the pockets appear, the scale ($\epsilon_F \neq 0$) associated with the size of the Fermi pockets is emergent. The long-range Coulomb interaction is screened by the Thomas-Fermi screening, and the Yukawa coupling induces the Landau damping term similar to the one of the Hertz-Millis theory. Thus, in spite of the presence of the gauge structure, the critical theory becomes
\begin{eqnarray}
\mathcal{S}_{H} &=& \int_{q} \Big(r+\gamma \frac{|q_n|}{|\vec{q}|} + |\vec{q}|^2\Big) | \mathcal{B}_i(q) |^2 + \frac{u}{4} \int_{x,\tau} (\mathcal{B}_i)^4 -h^i_{ext} \mathcal{B}_i.\nonumber
\end{eqnarray}
The term with $\gamma=\gamma(\epsilon_F)$ is for the Landau damping, and the coefficient of the term with $ |\vec{q}|^2$ is normalized to be one. We omit the term with $\mathcal{E}_i^2$ because it is irrelevant at the critical point (say, $r=0$).  
Namely, the dynamics of the gauge fluctuations are determined by the damping term, and the critical modes have the dynamical critical exponent $z_H=3$. Since  $d+z_H>4$, the term $(\mathcal{B}_i)^4$ with $u$ is irrelevant, and the gauge fluctuations are weakly correlated with $z_H=3$. The operator scaling dimensions are $[\mathcal{B}_i(x,\tau)]=2$, $[r]=2$, and $[h^i_{ext}]\equiv \nu_{ext}^{-1}=4$.

\textit{Multiscale quantum criticality} :  The interplay between the Kondo lattice coupling and the long-range Coulomb interaction naturally brings about Multiscale quantum criticality around the onset of $\langle B_i \rangle $. To see this, let us estimate the energy scale for breaking the particle-hole band condition by using Eq. \eqref{PH}. Setting $\Lambda \frac{d}{d \Lambda} \equiv \frac{d}{dl}$, the renormalization group equation is  $\frac{d}{d l} c_0 \simeq {0.3601}$, and the assumption $c_0 < c_1$ becomes invalid at $l^* \sim 3$. The associated energy scale is $E_{IR} \sim \Lambda^2 \,e^{-2 l^*} \sim \Lambda^2/ 400$ with the UV cutoff scale, $\Lambda$, below which the assumption of small $c_0$ breaks down. The energy scale $E_{IR}$ is much smaller than the band width of the conduction electron, which is of the order $\sim \Lambda^2$. It is natural to expect that the emergent particle-hole pocket-size scale ($\epsilon_F$) in the intermediate phase between F-LSM and F-FMSM is of similar order of magnitude, namely $E_{IR} \sim \epsilon_F$. 

Because of the hierarchy of energy scales, three sets of critical exponents would naturally appear in physical quantities near the onset of $\langle B_i \rangle $. 
For example, the emergent photons have $z_1=1$, the non-Fermi liquid excitations have $z_2 \sim 2$, and the Hertz-Millis fluctuations have $z_3=3$. The scaling dimensions of the external magnetic field are easily obtained by considering the coupling to the magnetic field. 
The emergent magnetic field couples to the Zeeman external magnetic field via $\int_{x,\tau} \vec{\mathcal{B}} \cdot \vec{h}_{ext}$, and the scaling dimension of the external field is $\nu_{ext,1}^{-1} =2$. The conduction electron couples to the external field as $\int_{x,\tau} \psi^{\dagger} M \psi \cdot \vec{h}_{ext}$, which gives $ \nu_{ext,2}^{-1} \sim 2$. 
We also show that the Hertz-Millis type fluctuation gives $\nu_{ext,3}^{-1} =4$.
 
Three different scaling behaviors can naturally arise in all physical quantities. For example, 
the magnetic Gruneissen parameter, $\Gamma_H = - (\partial M / \partial  T)_H /c_H$  
with magnetization ($M$) and  specific heat ($c_H$) at constant external magnetic field $H$, has the scaling form,
\begin{eqnarray}
\Gamma_H = \frac{1}{h_{ext}} \, \mathcal{F} (\frac{T^{1/(z_1 \nu_{ext,1})}}{h_{ext}}, \frac{T^{1/(z_2 \nu_{ext,2})}}{h_{ext}},  \frac{T^{1/(z_3 \nu_{ext,3})}}{h_{ext}}; \frac{T}{E_{IR}}). \nonumber
\end{eqnarray} 
The dimensionless function, $\mathcal{F}$, manifests the Multiscale quantum criticality. For example, when $T \gg E_{IR}$, one can find  $\mathcal{F}(x,y,z;0) = b_0+ b_1 x + b_2 y+b_3 z$ with three coefficients $b_{0,1,2,3}$ for $x,y,z \ll 1$. 

Possible exponents are $z_1 \nu_{ext,1} = 1/2$ for the emergent photons and $z_2 \nu_{ext,2}=1+O(1/N_f)$ for conduction electrons in F-LSM. 
Most importantly, near the quantum phase transition to the fractionalized ferromagnetic semi-metal state, the Hertz-Millis quantum critical point gives the scaling exponent, $z_3 \nu_{ext,3}= 3/4$. 
 It is interesting to note that similar multi scaling critical behavior in magnetic Gruneisen parameter is seen in Pr$_2$Ir$_2$O$_7$ \cite{Tokiwa3}.
Our theory naturally explains the appearance of Fermi pockets at low temperatures with multi scaling behaviors even though the calculated critical exponents are not exactly the same as the experimentally determined value. 

We also remark that our theory allows two channels, semimetallic conduction electrons and collective modes of the U(1) QSL, to contribute to magnetic susceptibility and other thermodynamic quantities. An interesting question is whether the contributions of such unusual excitations to thermodynamic and transport properties can explain various non-Fermi liquid behaviors seen in the experiment on Pr$_2$Ir$_2$O$_7$. We leave this intriguing problem for a future work. 


In conclusion, we investigate emergent quantum phenomena arising from the Kondo lattice coupling between the quantum spin liquid
of local moments and non-Fermi liquid conduction electrons in pyrochlore systems  $A_{2}B_{2}O_{7}$. 
Intertwined actions between the Kondo lattice coupling and the long-range Coulomb interaction are uncovered. 
As an important result, quantum criticality near the onset of ferromagnetic ordering naturally displays multi scaling behaviors.
Further works on more quantitative analysis and comparison with experiments are highly desired.  
 
\textit{Acknowledgement : }
This work was supported by NRF of Korea under Grant No. 2017R1C1B2009176 (HO, SL, EGM), 
and the NSERC of Canada, CIFAR, and Center for Quantum
Materials at the University of Toronto (YBK). 
This work was performed in part at the Aspen Center for Physics, 
which is supported by National Science Foundation grant PHY-1607611 (YBK). 
EGM also acknowledges the support of the POSCO Science Fellowship of POSCO TJ Park Foundation.

%

\onecolumngrid
\clearpage
\begin{center}
\textbf{\large Supplementary Information for ``Multiscale Quantum Criticality driven by Kondo lattice Coupling in Pyrochlore Systems''}
\end{center}
\begin{center}
{Hanbit Oh,$^1$ Sangjin Lee,$^1$ Yong Baek Kim,$^{2\textcolor{red}{*}}$ Eun-Gook Moon$^{1\textcolor{red}{\dagger}}$}\\
\emph{$^{1}$Department of Physics, Korea Advanced Institute of Science and Technology, Daejeon 305-701, Korea}\\
\emph{$^{2}$Department of Physics and Centre for Quantum Materials, University of Toronto, Toronto, Ontario M5S 1A7, Canada}
\end{center}
\setcounter{equation}{0}
\setcounter{figure}{0}
\setcounter{table}{0}
\setcounter{page}{1}
\setcounter{section}{0}
\setcounter{subsection}{0}

\maketitle 
\makeatletter
\renewcommand{\thesection}{\arabic{section}}
\renewcommand{\thesubsection}{\thesection.\arabic{subsection}}
\renewcommand{\thesubsubsection}{\thesubsection.\arabic{subsubsection}}
\renewcommand{\theequation}{S\arabic{equation}}
\renewcommand{\thefigure}{S\arabic{figure}}
\renewcommand{\thetable}{S\arabic{table}}

\section{Notation}
In this section, we provide detailed information on notations about symmetries and representations of our systems.
\vspace{-10pt}
\subsection{Fermionic Hamiltonian}
We consider a system with cubic and time reversal symmetries which realize a quadratic band touching energy spectrum in three spatial dimensions. Following the notation of [\onlinecite{Murakami}] , its low energy Hamiltonian, so-called Luttinger Hamiltonian, may be written as
\begin{eqnarray}
\mathcal{H}_{0}(\vec{k}) & = & \frac{c_{0}}{2m}\vec{k}^{2}+\sum_{a}\frac{\hat{c}_{a}}{2m}d_{a}(\vec{k})\Gamma_{a}, 
\end{eqnarray}
by introducing the four dimensional Gamma matrices ($\Gamma^{a}$) with $a=1,\cdots 5$. 
\begin{eqnarray}
\Gamma^{1}=\sigma^{z}\otimes \sigma^{y},& &\Gamma^{2}=\sigma^{z}\otimes \sigma^{x}, \nonumber\\
\Gamma^{3}=\sigma^{y}\otimes 1,\ & &\Gamma^{4}=\sigma^{x}\otimes 1,\nonumber\\
\Gamma^{5}=\sigma^{z}\otimes \sigma^{z}. & &
\end{eqnarray}
The Clifford algebra $\{\Gamma^{a},\Gamma^{b}\}=2\delta_{ab}$, is satisfied. The cubic symmetry gives the three independent parameters $c_0, c_1=\hat{c}_{1}=\hat{c}_{2}=\hat{c}_{3}, c_2=\hat{c}_{4}=\hat{c}_{5}$. The four band Hamiltonian may be expressed in terms of the $j=\frac{3}{2}$ angular momentum operators,
\begin{eqnarray}
\mathcal{H}_{0}(\vec{k})& = & \alpha_{1}k^{2}+\alpha_{2}(\vec{k}\cdot\vec{s})^{2}+\alpha_{3}(k_{x}^{2}s_{x}^{2}+k_{y}^{2}s_{y}^{2}+k_{z}^{2}s_{z}^{2})
\end{eqnarray}
with 
\begin{eqnarray}
s_{1} & = & \frac{\sqrt{3}}{2}\Gamma^{15}-\frac{1}{2}(\Gamma^{23}-\Gamma^{14}),\nonumber \\
s_{2} & = & -\frac{\sqrt{3}}{2}\Gamma^{25}+\frac{1}{2}(\Gamma^{13}+\Gamma^{24}),\nonumber \\
s_{3} & = & -\Gamma^{34}-\frac{1}{2}\Gamma^{12},
\end{eqnarray}

where $\Gamma^{ab}\equiv\frac{1}{2i}[\Gamma^{a},\Gamma^{b}]$ is used.
The bare femionic Green's function is defined as 
\begin{eqnarray}
G_{f}^{0}(k_{n},\vec{k})=(-ik_{n}+\mathcal{H}_{0}(\vec{k}))^{-1}=\sum_{\alpha=\pm}\frac{P_{\alpha}(\vec{k})}{-ik_{n}+E_{\alpha}(\vec{k})},
\end{eqnarray}
with $E_{\alpha}=\frac{c_{0}}{2m}\vec{k}^{2}+ \alpha E(\vec{k})=\frac{c_{0}}{2m}\vec{k}^{2}+  \frac{\alpha}{2m}\sqrt{\sum_a \hat{c}_a^2 d_a^2(\vec{k})}$. The projection operator, $P_{\alpha}(\vec{k})=\frac{1}{2}\left(1+\epsilon\frac{\mathcal{H}_{0}(\vec{k})-\frac{c_{0}}{2m}\vec{k}^{2}}{E(\vec{k})}\right)$ is introduced.
\vspace{20pt}

\subsection{Ferromagnetic and Paramagnetic states in spin-ice manifold}
\begin{figure}[H]
\begin{center}
\includegraphics[scale=1.2]{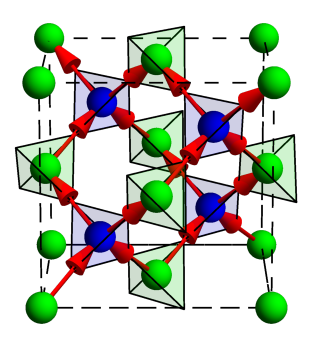}
\includegraphics[scale=1.2]{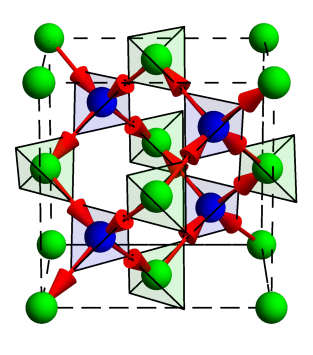}
\caption{Spin configurations of FM (left) and PM (right) states in spin-ice manifold.   The left figure has a nonzero value of $\left\langle \mathcal{B}(r)\right\rangle$  parallel to $\hat{z}$ axis. On the other hand, the right one has zero magnetization.}
\label{fig1}
\end{center} 
\end{figure}

%
The emergent magnetic field, $\mathcal{B}(v)$ on each spin site on a tetrahedra may be written as, 
\begin{eqnarray}
\mathcal{B}(r) & = & \sum_{i}\left\langle S_{i}(r)\right\rangle \mathbf{e}_{i} = (\mathcal{B}_{1}(r),\mathcal{B}_{2}(r),\mathcal{B}_{3}(r)),
\end{eqnarray}
where the summation is over four vertices of a tetrahedra. The unit vectors $\mathbf{e}_{i}$ are for four local easy axes given as $\mathbf{e}_{0}=\frac{1}{\sqrt{3}}(1,1,1),\ \mathbf{e}_{1}=\frac{1}{\sqrt{3}}(1,\overline{1},\overline{1}),\ \mathbf{e}_{2}=\frac{1}{\sqrt{3}}(\overline{1},1,\overline{1}),\ \mathbf{e}_{3}=\frac{1}{\sqrt{3}}(\overline{1},\overline{1},1)$. The spin-ice (2-in-2-out) manifold is defined as the divergence free condition, $\nabla \cdot \mathcal{B}=0$. \\
\indent Ferromagnetic (FM) and paramagnetic (PM) states in the spin-ice manifold are characterized by $\left\langle \mathcal{B}(r)\right\rangle \neq 0$ and $\left\langle \mathcal{B}(r)\right\rangle = 0$, respectively.  Typical configurations of the two states are illustrated in Fig.\ref{fig1}.  

\subsection{Action in Euclidean space}
\vspace{-5pt} 
One can derive the action in Euclidean space-time from Minkowski space-time by using analytic continuation, $e^{-i\mathcal{S}}\rightarrow e^{-\mathcal{S}_{E}}$.
%
\begin{eqnarray}
\mathcal{S} & = & \int_{\vec{x},t\ }\psi^{\dagger}\left(\alpha i\partial_{t}-\mathcal{H}_{0}(-i\nabla)\right)\psi+\frac{\mathcal{E}^{2}}{2}-\frac{\mathcal{B}^{2}}{2}-g_{E}\psi^{\dagger}\hat{M}\psi\mathcal{\cdot\vec{E}}-g_{B}\psi^{\dagger}\hat{M}_{B}\psi\mathcal{\cdot\vec{B}}-ie\psi^{\dagger}\psi \varphi,\\
 & = & \int_{\vec{k},\omega\ }\psi_{k}^{\dagger}\left(\alpha \omega-\mathcal{H}_{0}(\vec{k})\right)\psi_{k}+\frac{1}{2}\mathcal{E}_{k}^{2}-\frac{1}{2}\mathcal{B}_{k}^{2}-\int_{\begin{subarray}{l}\vec{k},\omega\\\vec{q},\Omega\end{subarray}} g_{E}\psi_{k}^{\dagger}\hat{M}_{E}(\vec{k},\vec{q})\psi_{k+q}\mathcal{\cdot\vec{E}}_{q}+g_{B}\psi_{k}^{\dagger}\hat{M}_{B}(\vec{k},\vec{q})\psi_{k+q}\mathcal{\cdot\vec{B}}_{q}\nonumber\\
&&\ \ \ \ \ \ \ \ +ie\psi_{k}^{\dagger}\hat{M}_{\varphi}(\vec{k},\vec{q})\psi_{k+q}\varphi_{q},
\label{MA}
 \end{eqnarray}
 \vspace{-10pt}
 \begin{eqnarray}
\mathcal{S}_{E} & = & \int_{\vec{x},\tau\ }\psi^{\dagger}\left(\alpha\partial_{\tau}+\mathcal{H}_{0}(-i\nabla)\right)\psi-\frac{\mathcal{E}^{2}}{2}+\frac{\mathcal{B}^{2}}{2}+g_{E}\psi^{\dagger}\hat{M}_{E}\psi\mathcal{\cdot\vec{E}}+g_{B}\psi^{\dagger}\hat{M}_{B}\psi\mathcal{\cdot\vec{B}},\\
 & = & \int_{\vec{k},k_{n}\ }\psi_{k_{E}}^{\dagger}\left(-\alpha ik_{n}+\mathcal{H}_{0}(\vec{k})\right)\psi_{k_{E}}-\frac{1}{2}\mathcal{E}_{k_{E}}^{2}+\frac{1}{2}\mathcal{B}_{k_{E}}^{2}+\int_{\begin{subarray}{l}\vec{k}k_{n}\\ \vec{q},k_{n}\end
{subarray}}g_{E}\psi_{k_{E}}^{\dagger}\hat{M}_{E}(\vec{k},\vec{q})\psi_{k+q}+g_{B}\psi_{k_{E}}^{\dagger}\hat{M}_{B}(\vec{k},\vec{q})\psi_{k_{E}+q_{E}}\mathcal{\cdot\vec{B}}_{q_{E}}\nonumber\\
&&\ \ \ \ \ \ \ \ +ie\psi_{k_{E}}^{\dagger}\hat{M}_{\varphi}(\vec{k},\vec{q})\psi_{k_{E}+q_{E}}\varphi_{q_{E}}.
\label{EA}
\end{eqnarray}
where $\int_{\vec{k},\omega}=\int \frac{d^{3}k d\omega}{(2\pi)^{4}}$, $\int_{\vec{k},k_{n}}=\int \frac{d^{3}k dk_{n}}{(2\pi)^{4}}$ and $\hat{M}_{\varphi}(\vec{k},\vec{q})=I$ are used. Vertex functions with emergent gauge field, $\hat{M}_{B,E}(\vec{k},\vec{q})$ are summarized in the next section. One can find the minus sign in front of the $\vec{\mathcal{E}}$ in $\mathcal{S}_{E}$. We will omit the subscript $E$ of Euclidean variables for simplicity. 

Notations for physical quantities in the two space-times are summarized in Table \ref{table1}.\\
\begin{table}[H]
\begin{center}
\renewcommand{\arraystretch}{1.4}
\begin{tabular}{c|c}
\hline 
\hline
Minkowski & Euclidean\tabularnewline
\hline 
$x^{\mu}\equiv(t,\vec{x})$ & $x_{E}^{\mu}\equiv(\tau=it,\vec{x})$\\ 
$k^{\mu}\equiv(\omega,\vec{k})$ & $k_{E}^{\mu}\equiv(k_{n},\vec{k})$\\
$\mathcal{A}^{\mu}\equiv(\phi,\vec{A})$ & $\mathcal{A}_{E}^{\mu}\equiv(i\phi,\vec{A})$\\
$F^{\mu\nu}\equiv\partial^{\mu}A^{\nu}-\partial^{\nu}A^{\mu}$ & $F_{E}^{\mu\nu}\equiv\partial_{E}^{\mu}A_{E}^{\nu}-\partial_{E}^{\nu}A_{E}^{\mu}$\\
$\mathcal{E}^{i}\equiv F^{0i},\ \mathcal{B}^{i}\equiv-\epsilon^{ijk}F^{jk}$ & $i\mathcal{E}^{i}\equiv F_{E}^{0i},\ \mathcal{B}^{i}\equiv-\epsilon^{ijk}F_{E}^{jk}$\tabularnewline
\hline 
\hline
\end{tabular}
\caption{Variables  in Minkowskoi and Euclidean space-times. 
}
\label{table1}
\end{center}
\end{table}

\subsection{The couplings between fermions and emergent gauge fields}
Here we construct the Yukawa couplings between conduction electrons and gauge fields by inspecting the symmetries of the system. The key symmetries are time reversal (TR), inversion (Inv) and $O_{h}$ symmetry. Since $\vec{\mathcal{B}},\ \vec{\mathcal{E}}$ are vectors, their coupling matrices may couple to $T_{1g}$ irreducible representations (Irreps) of $O_{h}$ which transform $\hat{M}_{B}\rightarrow -\hat{M}_{B}$, $\hat{M}_{E}\rightarrow \hat{M}_{E}$ under TR and $\hat{M}_{B}\rightarrow\hat{M}_{B}$, $\hat{M}_{E}\rightarrow-\hat{M}_{E}$ under Inv.\\
\indent The symmetries of Gamma matrices are well-known, and it is easy to check that $\Gamma_{a}$ matrices are even under both Inv and TR, while $\Gamma^{ab}$ matrices are even under Inv but odd under TR. 
Under $O_{h}$ operations, $\Gamma^{a}$ split into $T_{2g}+ E_{g}$ and $\Gamma^{a,b}$ split into $A_{2g}+2\ T_{1g}+\ T_{2g}$. \\
\indent The Yukawa coupling matrices $\hat{M}_{B},\ \hat{M}_{E}$ are summarized in Fig.\ref{fig2}. The emergent magnetic field $\vec{\mathcal{B}}$ couple to fermion with $\vec{s}$ and  $\vec{\tilde{s}}\equiv -\frac{41}{12}\vec{s}+\frac{5}{3}(s_{x}^{3},s_{y}^{3},s_{z}^{3})$ in the lowest order. 
Similarly, the emergent electric field $\vec{\mathcal{E}}$ couples to fermion with $\nabla,\ \nabla \times \vec{s},\ \nabla \times \vec{\tilde{s}}$. 
Engineering dimensions of the coupling constants are easily evaluated and we conclude that $g_{B}$ is more relevant than $g_{E}$ at low energies. Thus, we concentrate on the coupling $g_{B}$ and omit the subscript $B$ in the main text.

\begin{figure}[H] 
\begin{center}
\begin{tikzpicture}[anchor=base,baseline]

\begin{scope}[scale={1}]

\draw[thick,decorate,decoration={complete sines, number of sines=4,amplitude=5pt}](0,0)--(-1,0);
\node[left] at(-1,0.0) {$\vec{\mathcal{B}}_{\vec{q}}$};
\node[right] at(0.3,0.0) {$=g_{B}\vec{s},\ g_{B}\vec{\tilde{s}}\ \cdots $};
\node[left] at(0.1,-0.6) {$\vec{k}$};
\node[left] at(0.2,0.6) {$\vec{k}+\vec{q}$};

\draw[thick,postaction={decorate},decoration={markings, mark=at position 0.6 with {\arrow{latex}}}](0.0,0.0)--(0.5,0.8)node[pos=0.52,above]{};
\draw[thick,postaction={decorate},decoration={markings, mark=at position 0.6 with {\arrow{latex}}}](0.5,-0.8)--(0,0)node[pos=0.52,left]{};

\end{scope}
\end{tikzpicture}
\begin{tikzpicture}[anchor=base,baseline]

\begin{scope}[scale={1}]

\draw[thick,decorate,decoration={complete sines, number of sines=4,amplitude=5pt}](0,0)--(-1,0);
\node[left] at(-1,0.0) {$\vec{\mathcal{E}}_{\vec{q}}$};
\node[right] at(0.3,0.0) {$=ig_{E_{1}}\vec{q},\ g_{E}(2\vec{k}+\vec{q})\times\vec{s},\ g_{E}(2\vec{k}+\vec{q})\times \vec{\tilde{s}}\cdots$};
\node[left] at(0.1,-0.6) {$\vec{k}$};
\node[left] at(0.2,0.6) {$\vec{k}+\vec{q}$};

\draw[thick,postaction={decorate},decoration={markings, mark=at position 0.6 with {\arrow{latex}}}](0.0,0.0)--(0.5,0.8)node[pos=0.52,above]{};
\draw[thick,postaction={decorate},decoration={markings, mark=at position 0.6 with {\arrow{latex}}}](0.5,-0.8)--(0,0)node[pos=0.52,left]{};

\end{scope}
\end{tikzpicture}
\caption{Diagrammatic expressions of the Yukawa coupling matrices, $\hat{M}_{B}(\vec{k},\vec{q})$ and $\hat{M}_{E}(\vec{k},\vec{q})$. }
\label{fig2}
\end{center}
\end{figure}
\vspace{-15pt}

\vspace{15pt}
\section{Perturbative Kondo lattice coupling regime}
In this section, we consider the perturbative regime 
and show that the PHS breaking term is generated by gauge fluctuation even for $g_{B}\ll g_{B,c}$. 
We use the bare Green's function with the transverse condition to calculate the fermionic self-energy by gauge fluctuation.
\begin{eqnarray}
G_{B,T}^{ij}(k_{n},\vec{k})& \simeq &\mu_{0}(\delta^{ij}-\hat{k}^{i}\hat{k}^{j}).
\end{eqnarray}
The PHS breaking term from the gauge fluctuations has linear dependence on UV cutoff, $\Lambda$ instead of logarithmic dependence.
\begin{eqnarray}
\Sigma^{0}_{f}(q_{n},\vec{q}) & = &g_{B}^{2}\mu_{0}\Lambda\left( \delta c_{0} \frac{\vec{q}^{2}}{2m}\right),
\end{eqnarray}
where $\delta c_{0}=+0.025$. The structure of coupling vertex and the transverse condition are essential ingredients for the nonzero PHS breaking term. We summarize the results in Table.\ref{table3}. 

Note that the PHS is realized with the long-range Coulomb interaction since the Coulomb interaction dominates over Kondo lattice coupling in the weak coupling regime.
\begin{table}[H]
\begin{center}
\renewcommand{\arraystretch}{1.5}
\begin{tabular}{C{0.8cm} |C{0.8cm} |C{1.4cm}| C{1.4cm}}
\hline 
\hline 
$R$ & $\hat{M}$& $\delta c_{0,B}$ & $\delta c_{B,T}$\tabularnewline
\hline
$A_{2}$ & $\Gamma^{45}$ & $0$ & $0$\tabularnewline
\hline 
\multirow{2}{*}{$T_{1}$} & $\vec{s}$ & $0$ &  $+0.025$
\tabularnewline
\cline{2-4} 
 & $\vec{\tilde{s}}$ & $0$ &  $-0.006$\tabularnewline
\hline 
$T_{2}$ & $\vec{T}$ & $0$ &  $+0.019$\tabularnewline
\hline 
\hline 
\end{tabular}\caption{The PHS breaking parts of the fermionic self-energy by fluctuation of TRS odd order parameters. $\delta c_{0}$ depends on the matrix structure of the vertex and the transverse condition. $\delta c_{0,B}$ ($\delta c_{0,T}$) is the result from the bare (transverse) part of bosonic Green's function, $G_{B}^{ij}\propto \delta^{ij}$ ($G_{B,T}^{ij}\propto (\delta^{ij}-\hat{k}^{i}\hat{k}^{j})$). The coupling with $A_{2}$ order parameter does not generate $\delta c_{0,B},\delta c_{0,T}$ regardless of imposition of transverse condition. However the couplings with $T_{1},\ T_{2}$ order parameters generate $\delta c_{0,T}$ only when we impose the transverse condition. We use $\vec{\tilde{s}}\equiv -\frac{41}{12}\vec{s}+\frac{5}{3}(s_{x}^{3},s_{y}^{3},s_{z}^{3})$ and $\vec{T}=\sqrt{\frac{5}{12}}\left( \left\lbrace s_{x},s_{y}^{2}-s_{z}^{2}\right\rbrace, \left\lbrace s_{y},s_{z}^{2}-s_{x}^{2}\right\rbrace, \left\lbrace s_{z},s_{x}^{2}-s_{y}^{2}\right\rbrace \right)$ .}
\vspace{-10pt}
\label{table3}
\end{center}
\end{table}
\vspace{-15pt}
\section{STABILITY OF Gauge fluctuation near the critical point}
The bosonic self-energy $\Sigma_{\mathcal{B},T}(\vec{q},q_{n})$ is not universal but dependent on microscopic details of conducting fermions and background spin configurations. We provide the details of the bosonic self-energy at the one-loop level by controlling $c_{0},c_{1},c_{2}$ and the transverse condition of the gauge field. 
 \vspace{-10pt}
\subsection{The most symmetric case ($c_{0}=0$, $c_{1}=c_{2}=1$)}
 \vspace{-5pt}
Here we mainly consider the most symmetric condition with SO(3) and PHS symmetries with $\nabla \cdot \mathcal{B}=0$ calculating the bosonic self-energy of nonzero external momentum and frequency. We obtain $\delta\Sigma_{\mathcal{B},T}^{ij}(\vec{q},q_{n})$ by subtracting the UV divergent piece $\Sigma_{\mathcal{B}}^{ij}(0,0)=-\frac{1}{2\pi^{2}}g_{B}^{2}\Lambda\delta^{ij}$.
\begin{eqnarray}
\delta\Sigma_{\mathcal{B},T}^{ij}(\vec{q},q_{n}) & = &P_{T}^{il}(\vec{q}) \left(\Sigma_{\mathcal{B}}^{lk}(\vec{q},q_{n})-\Sigma_{\mathcal{B}}^{lk}(0,0)\right)P_{T}^{kj}(\vec{q}),\nonumber\\
 & = & g_{B}^{2}\left(a_{T}\left|\vec{q}\right|+a_{\omega}\sqrt{\left|q_{n}\right|}\right)P_{T}^{ij}(\vec{q}),
\end{eqnarray}
where the dimensionless coefficients $a_{T}=+0.036,a_{\omega}=+0.040$ and the projection operator $P_{T}^{ij}(\vec{q})=(\delta^{ij}-\hat{q}^{i}\hat{q}^{j})$ are used.
\vspace{-15pt}
\subsection{The case with broken particle-hole symmetry ($c_{0}\neq0$, $c_{1}=c_{2}=1$)} 
\begin{figure}[H]
\begin{center}
\includegraphics[scale=0.5]{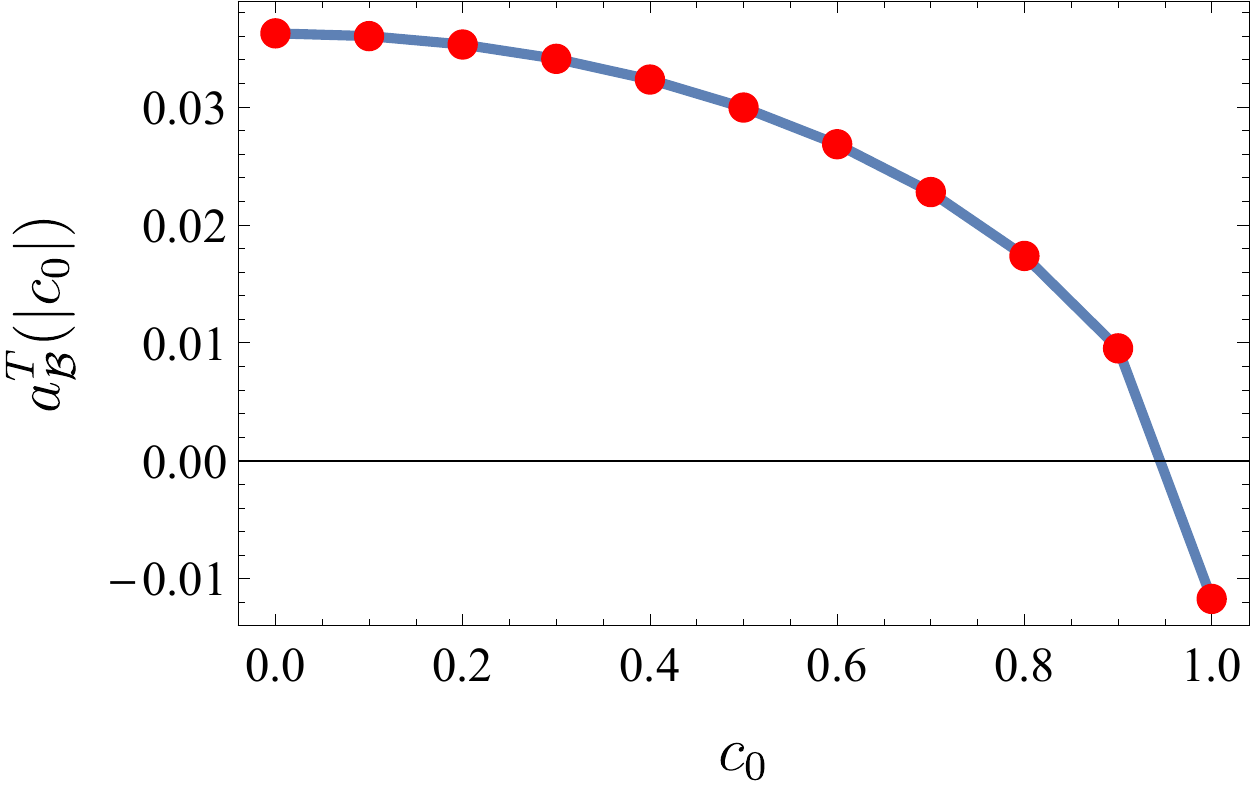}\ \ \includegraphics[scale=0.5]{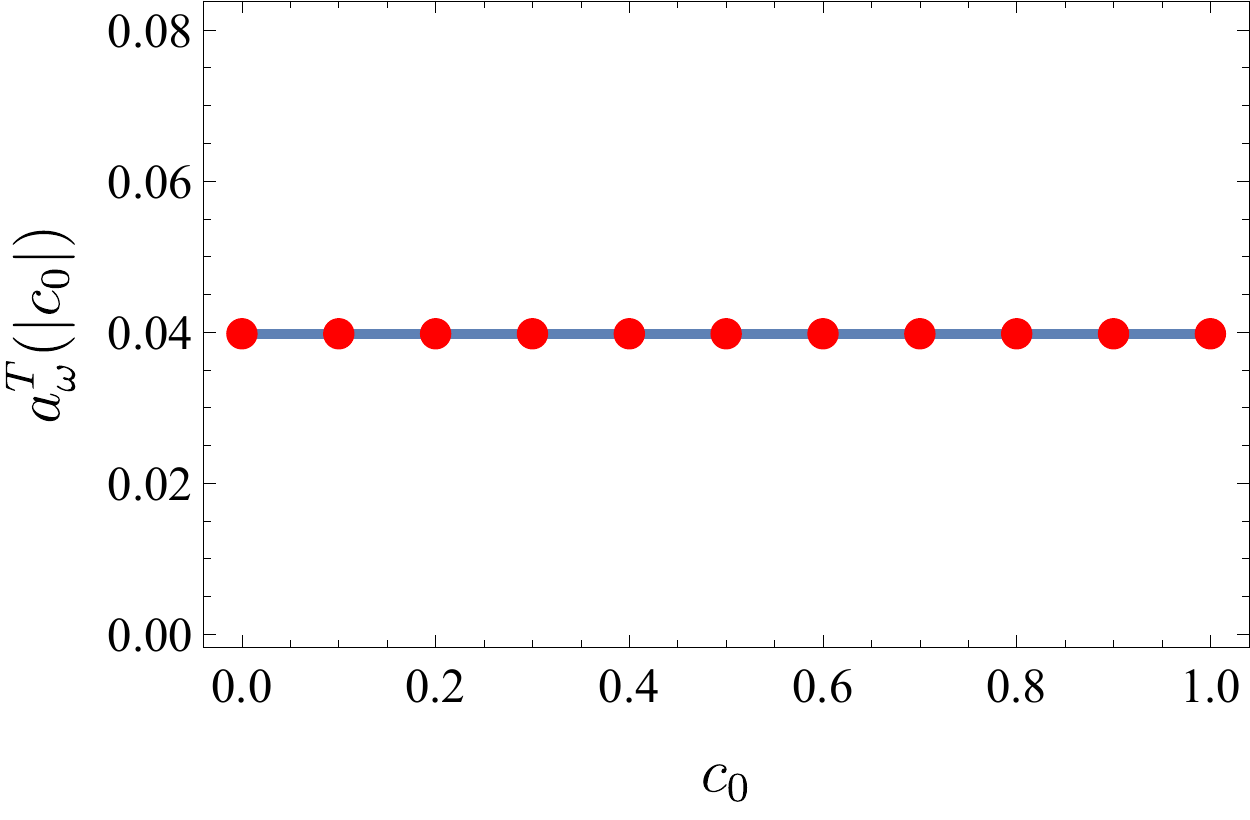}
\vspace{-10pt}
\caption{The dimensionless function $a_{T}(\left|c_{0}\right|),\ a_{\omega}(\left|c_{0}\right|)$. }

\vspace{-15pt}
\label{fig3}
\end{center}
\end{figure}

 \vspace{-5pt}
Here we consider the case with SO(3) symmetry and $\nabla \cdot \mathcal{B}=0$, but in the absence of PHS symmetry. We obtain $\delta\Sigma_{\mathcal{B},T}^{ij}(\vec{q},q_{n};c_{0})$ by subtracting the UV divergent piece $\Sigma_{\mathcal{B}}^{ij}(0,0;c_{0})=-\frac{1}{2\pi^{2}}g_{B}^{2}\Lambda\delta^{ij}$.
\begin{eqnarray}
\delta\Sigma_{\mathcal{B},T}^{ij}(\vec{q},q_{n};c_{0}) & = &P_{T}^{i,l}(\vec{q}) \left( \Sigma_{\mathcal{B}}^{lk}(\vec{q},q_{n};c_{0})-\Sigma_{\mathcal{B}}^{lk}(0,0)\right)P_{T}^{k,j}(\vec{q}),\nonumber\\
 & = & g_{B}^{2}\left(a_{T}(c_{0})\left|\vec{q}\right|+a_{\omega}(c_{0})\sqrt{\left|q_{n}\right|}\right)P_{T}^{ij}(\vec{q}),
\end{eqnarray}
where the dimensionless coefficients of external momentum and frequency dependence $a_{T}(c_{0}),\ a_{\omega}(c_{0})$ are given in Fig. \ref{fig3} and the projection operator $P_{T}^{ij}(\vec{q})=(\delta^{ij}-\hat{q}^{i}\hat{q}^{j})$ is used. $a_{T}(c_{0})$ changes and even becomes negative in the $c_{0}\rightarrow1$ limit, whereas $a_{\omega}(c_{0})$ is constant as varying $c_{0}$. 

\subsection{The case with broken rotational symmetry($c_{0}=0$, $c_{1}\neq c_{2}$)}
\begin{figure}[H]
\begin{center}
\includegraphics[scale=0.56]{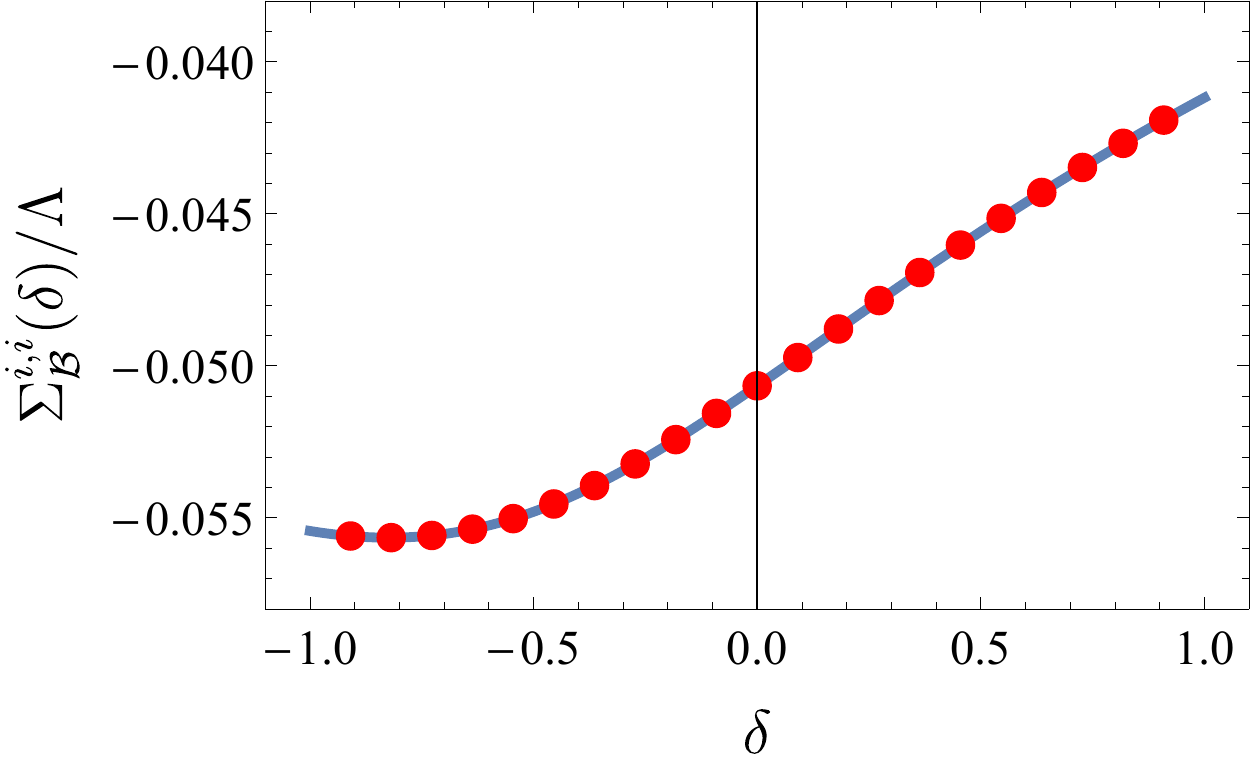}\ \ \ \ 
\includegraphics[scale=0.38]{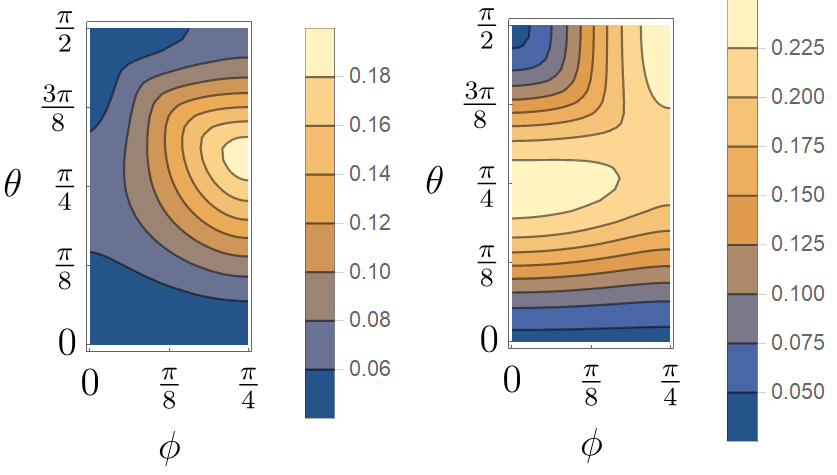}
\vspace{-5pt}
\caption{The UV divengent piece $\Sigma_{\mathcal{B}}^{ii}(0,0;\delta)$ and the dimensionless function $\ a_{T,1}^{T,(1)}(\hat{q}),\ a_{T,2}^{T,(1)}(\hat{q})$.}
\label{fig4}
\end{center}
\end{figure}

Here we consider the case with PHS symmetry and $\nabla \cdot \mathcal{B}=0$, but in the absence of SO(3) symmetry. We control the relative amplitude of coefficients between $T_{2g}$ and $E_{g}$ channel, called $c_{1},c_{2}$.  We obtain the finite part $\Sigma_{\mathcal{B}}^{ij}(0,0;c_{1},c_{2})$ by subtracting the UV divergent part  $\Sigma_{\mathcal{B}}^{ij}(0,0;c_{1},c_{2})$ whic is a function $c_{1}$ and $c_{2}$ as in Fig. \ref{fig4}.
\begin{eqnarray}
\delta\Sigma_{\mathcal{B},T}^{ij}(\vec{q},q_{n};c_{1},c_{2}) & = & P_{T}^{il} (\vec{q})\left( \Sigma_{\mathcal{B}}^{lk}(\vec{q},q_{n};c_{1},c_{2})-\Sigma_{\mathcal{B}}^{lk}(0,0;c_{1},c_{2})\right)P_{T}^{kj} (\vec{q}),\nonumber\\
 & = & g_{B}^{2}\sum_{\alpha=1,2}\left(a_{T,\alpha}(\hat{q};c_{1},c_{2})\left|\vec{q}\right|+a_{\omega}(c_{1},c_{2})\sqrt{\left|q_{n}\right|}\right)v_{T,\alpha}^{i}(\hat{q})v_{T,\alpha}^{j}(\hat{q}),
\end{eqnarray}
where $a_{T,\alpha=1,2}^{T}(\hat{q};c_{1},c_{2})$ are the two different transverse modes of $\delta\Sigma_{\mathcal{B},T}^{ij}(\vec{q},0;c_{1},c_{2})$ and $v_{T,\alpha}^{i}(\hat{q})$ are $i$-th component of conjugate eigenvectors. Without SO(3) symmetry, the two transverse modes are not degenerate but split into two different modes. If we treat the degree of anisotropy $\delta=\frac{c_{1}-c_{2}}{c_{1}+c_{2}}$ as perturbation, $ a_{T,\alpha}$ and $a_{\omega}$ are expanded from the isotropic case results.
\begin{eqnarray}
a_{T,\alpha}^{T}(\hat{q};\delta)&\simeq &a_{T}^{(0)}+\delta\times a_{T,\alpha}^{(1)}(\hat{q}),\nonumber \\
a_{\omega}(\delta)&\simeq&a_{\omega}^{(0)}+\delta \times a_{\omega}^{(1)},\nonumber
\end{eqnarray}
 where $a_{T}^{(0)}=+0.036,\ a_{\omega}^{(0)}=+0.040$, $a_{\omega}^{(1)}=-0.007$ and $a_{T,\alpha}^{(1)}(\hat{q})$ are given in Fig.\ref{fig4}. There are several things to remark. First, $\Sigma_{\mathcal{B}}^{ii}(0,0;\delta)$ is always negative value in the $-1<\delta<1$ range. Second, we can take finite range of $\delta$  that both  $a_{T}(\hat{q};\delta)$  and $a_{\omega}(\delta)$ are positive. 
\vspace{-10pt}

\vspace{10pt}
\subsection{Out of spin-ice manifold} 
Here we consider spin configurations out of the spin-ice manifold, which release the divergence free condition $\nabla \cdot \mathcal{B}=\rho_{M}\neq 0$ in our effective theory. The key difference here is to consider not only the transverse mode but also the longitudinal mode of the bosonic self-energy. In the presence of SO(3) and PHS symmetries,
\begin{eqnarray}
\delta\Sigma_{\mathcal{B}}^{ij}(\vec{q},q_{n})^{-1} &=& \left[\frac{P^{ij}_{T}(\vec{q})}{g_{B}^{2}\left(a_{T}\left|\vec{q}\right|+a_{\omega}\sqrt{\left|q_{n}\right|}\right)}+\frac{P^{ij}_{L}(\vec{q})}{g_{B}^{2}\left(a_{L}\left|\vec{q}\right|+a_{\omega}\sqrt{\left|q_{n}\right|}\right)}\right],
\end{eqnarray}
where the dimensionless parameters $a_{T}=0.036,\ a_{L}=-0.004,\ a_{\omega}=0.040$ are given in the previous section and the projection operators $P_{T}^{ij}(\vec{q})=\left(\delta^{ij}- \hat{q}^{i}\hat{q}^{j}\right)$, $P_{L}^{ij}(\vec{q})=\hat{q}^{i}\hat{q}^{j}$ are used.\\
\indent Thus, we conclude that the quantum phase transitions  and their scenarios of other spin configurations may be different.

\section{Large $N_{f}$ Calculation} 
Here we provide the calculation details of the large $N_{f}$ analysis. We show that the fixed point with $c_{0}=\alpha=\delta=0$ point is unstable where $\alpha$ is for the ratio between $\vec{s}$ and $\vec{\tilde{s}}$, and $\delta$ is for the degree of anisotropy. We first calculate the divergent PHS breaking term in the fermionic self-energy, so $c_0 \neq 0$. Second, we find that $\left[s\right]>\left[\tilde{s}\right]$ which is not consistent with the original condition indicating $\alpha \neq 0$. 
\subsection{Green's function and self-energy conventions}
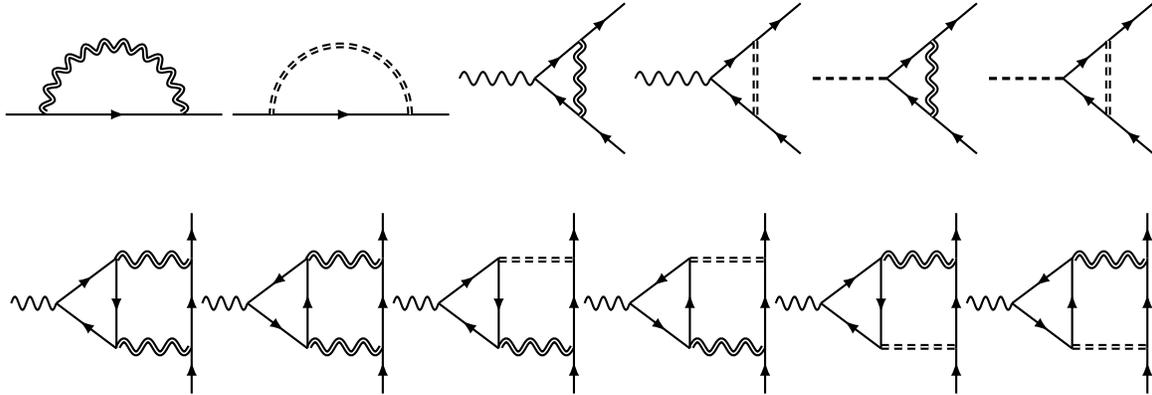
\begin{figure}[H]
\begin{center}
\begin{tikzpicture}[anchor=base,baseline]
\begin{scope}[scale={0.8}]
\draw [double,thick] (0,-1.69) to [wavy semicircle, wave amplitude=0.085cm,wave count=11] (0,0.55);
\draw[thick,postaction={decorate},decoration={markings, mark=at position 0.55 with {\arrow{latex}}}](-1.8,-0.6)--(1.8,-0.6);              
\end{scope}
\end{tikzpicture}
\begin{tikzpicture}[anchor=base,baseline]
\begin{scope}[scale={0.8}]
\draw[thick,postaction={decorate},decoration={markings, mark=at position 0.55 with {\arrow{latex}}}](-1.8,-0.6)--(1.8,-0.6);              
\draw[thick,double,densely dashed, black,postaction={decorate}](1.15,-0.6) arc (0:180:1.15cm);
\end{scope}
\end{tikzpicture} 
\begin{tikzpicture}[anchor=base,baseline]
\begin{scope}[scale={1}]
\draw[thick,decorate,decoration={complete sines, number of sines=4,amplitude=5pt}](0,0)--(-1,0);
\draw[thick,postaction={decorate},decoration={markings, mark=at position 0.6 with {\arrow{latex}}}](0.0,0.0)--(0.6,0.5)node[pos=0.52,above]{};
\draw[thick,postaction={decorate},decoration={markings, mark=at position 0.6 with {\arrow{latex}}}](0.6,-0.5)--(0,0)node[pos=0.52,left]{};
\draw[thick,postaction={decorate},decoration={markings, mark=at position 0.6 with {\arrow{latex}}}](0.6,0.5)--(1.2,1)node[pos=0.52,above]{};
\draw[thick,postaction={decorate},decoration={markings, mark=at position 0.6 with {\arrow{latex}}}](1.2,-1.0)--(0.6,-0.5)node[pos=0.52,left]{};
\draw[double,thick,decorate,decoration={complete sines, number of sines=3.5,amplitude=3pt}](0.6,0.5)--(0.6,-0.5);
\end{scope}
\end{tikzpicture}\ 
\begin{tikzpicture}[anchor=base,baseline]
\begin{scope}[scale={1}]
\draw[thick,decorate,decoration={complete sines, number of sines=4,amplitude=5pt}](0,0)--(-1,0);
\draw[thick,postaction={decorate},decoration={markings, mark=at position 0.6 with {\arrow{latex}}}](0.0,0.0)--(0.6,0.5)node[pos=0.52,above]{};
\draw[thick,postaction={decorate},decoration={markings, mark=at position 0.6 with {\arrow{latex}}}](0.6,-0.5)--(0,0)node[pos=0.52,left]{};
\draw[thick,postaction={decorate},decoration={markings, mark=at position 0.6 with {\arrow{latex}}}](0.6,0.5)--(1.2,1)node[pos=0.52,above]{};
\draw[thick,postaction={decorate},decoration={markings, mark=at position 0.6 with {\arrow{latex}}}](1.2,-1.0)--(0.6,-0.5)node[pos=0.52,left]{};
\draw[double,thick,decorate,densely dashed](0.6,0.5)--(0.6,-0.5);
\end{scope}
\end{tikzpicture}\ 
\begin{tikzpicture}[anchor=base,baseline]
\begin{scope}[scale={1}]
\draw[very thick,decorate,densely dashed](0,0)--(-1,0);
\draw[thick,postaction={decorate},decoration={markings, mark=at position 0.6 with {\arrow{latex}}}](0.0,0.0)--(0.6,0.5)node[pos=0.52,above]{};
\draw[thick,postaction={decorate},decoration={markings, mark=at position 0.6 with {\arrow{latex}}}](0.6,-0.5)--(0,0)node[pos=0.52,left]{};
\draw[thick,postaction={decorate},decoration={markings, mark=at position 0.6 with {\arrow{latex}}}](0.6,0.5)--(1.2,1)node[pos=0.52,above]{};
\draw[thick,postaction={decorate},decoration={markings, mark=at position 0.6 with {\arrow{latex}}}](1.2,-1.0)--(0.6,-0.5)node[pos=0.52,left]{};
\draw[double,thick,decorate,decoration={complete sines, number of sines=3.5,amplitude=3pt}](0.6,0.5)--(0.6,-0.5);
\end{scope}
\end{tikzpicture}\ 
\begin{tikzpicture}[anchor=base,baseline]
\begin{scope}[scale={1}]
\draw[very thick,decorate,densely dashed](0,0)--(-1,0);
\draw[thick,postaction={decorate},decoration={markings, mark=at position 0.6 with {\arrow{latex}}}](0.0,0.0)--(0.6,0.5)node[pos=0.52,above]{};
\draw[thick,postaction={decorate},decoration={markings, mark=at position 0.6 with {\arrow{latex}}}](0.6,-0.5)--(0,0)node[pos=0.52,left]{};
\draw[thick,postaction={decorate},decoration={markings, mark=at position 0.6 with {\arrow{latex}}}](0.6,0.5)--(1.2,1)node[pos=0.52,above]{};
\draw[thick,postaction={decorate},decoration={markings, mark=at position 0.6 with {\arrow{latex}}}](1.2,-1.0)--(0.6,-0.5)node[pos=0.52,left]{};
\draw[double,thick,decorate,densely dashed](0.6,0.5)--(0.6,-0.5);
\end{scope}
\end{tikzpicture}
\end{center}
\begin{center}
\begin{tikzpicture}[anchor=base,baseline]
\begin{scope}[scale={1}]
\draw[thick,decorate,decoration={complete sines, number of sines=3,amplitude=5pt}](0,0)--(-0.6,0);
\draw[thick,postaction={decorate},decoration={markings, mark=at position 0.6 with {\arrow{latex}}}](0.0,0.0)--(0.8,0.6)node[pos=0.52,above]{};
\draw[thick,postaction={decorate},decoration={markings, mark=at position 0.6 with {\arrow{latex}}}](0.8,-0.6)--(0,0)node[pos=0.52,left]{};
\draw[double,thick,decorate,decoration={complete sines, number of sines=3,amplitude=5pt}](0.82,-0.58)--(1.78,-0.58);
\draw[double,thick,decorate,decoration={complete sines, number of sines=3,amplitude=5pt}](0.8,0.58)--(1.78,0.58);
\draw[thick,postaction={decorate},decoration={markings, mark=at position 0.6 with {\arrow{latex}}}](0.8,0.6)--(0.8,-0.6);
\draw[thick,postaction={decorate},decoration={markings, mark=at position 0.7 with {\arrow{latex}}}](1.8,-1.2)--(1.8,-0.6);
\draw[thick,postaction={decorate},decoration={markings, mark=at position 0.6 with {\arrow{latex}}}](1.8,-0.6)--(1.8,0.6);
\draw[thick,postaction={decorate},decoration={markings, mark=at position 0.7 with {\arrow{latex}}}](1.8,0.6)--(1.8,1.2);
\end{scope}
\end{tikzpicture}
\begin{tikzpicture}[anchor=base,baseline]
\begin{scope}[scale={1}]
\draw[thick,decorate,decoration={complete sines, number of sines=3,amplitude=5pt}](0,0)--(-0.6,0);
\draw[thick,postaction={decorate},decoration={markings, mark=at position 0.6 with {\arrow{latex}}}](0.80,0.60)--(0.0,0.0)node[pos=0.52,above]{};
\draw[thick,postaction={decorate},decoration={markings, mark=at position 0.6 with {\arrow{latex}}}](0,0)--(0.8,-0.6)node[pos=0.52,right]{};
\draw[double,thick,decorate,decoration={complete sines, number of sines=3,amplitude=5pt}](0.82,-0.58)--(1.78,-0.58);
\draw[double,thick,decorate,decoration={complete sines, number of sines=3,amplitude=5pt}](0.8,0.58)--(1.78,0.58);
\draw[thick,postaction={decorate},decoration={markings, mark=at position 0.6 with {\arrow{latex}}}](0.8,-0.6)--(0.8,0.6);
\draw[thick,postaction={decorate},decoration={markings, mark=at position 0.7 with {\arrow{latex}}}](1.8,-1.2)--(1.8,-0.6);
\draw[thick,postaction={decorate},decoration={markings, mark=at position 0.6 with {\arrow{latex}}}](1.8,-0.6)--(1.8,0.6);
\draw[thick,postaction={decorate},decoration={markings, mark=at position 0.7 with {\arrow{latex}}}](1.8,0.6)--(1.8,1.2);
\end{scope}
\end{tikzpicture}
\begin{tikzpicture}[anchor=base,baseline]
\begin{scope}[scale={1}]
\draw[thick,decorate,decoration={complete sines, number of sines=3,amplitude=5pt}](0,0)--(-0.6,0);
\draw[thick,postaction={decorate},decoration={markings, mark=at position 0.6 with {\arrow{latex}}}](0.0,0.0)--(0.8,0.6)node[pos=0.52,above]{};
\draw[thick,postaction={decorate},decoration={markings, mark=at position 0.6 with {\arrow{latex}}}](0.8,-0.6)--(0,0)node[pos=0.52,left]{};
\draw[double,thick,decorate,decoration={complete sines, number of sines=3,amplitude=5pt}](0.82,-0.58)--(1.78,-0.58);
\draw[double,thick,decorate,densely dashed](0.8,0.58)--(1.78,0.58);
\draw[thick,postaction={decorate},decoration={markings, mark=at position 0.6 with {\arrow{latex}}}](0.8,0.6)--(0.8,-0.6);
\draw[thick,postaction={decorate},decoration={markings, mark=at position 0.7 with {\arrow{latex}}}](1.8,-1.2)--(1.8,-0.6);
\draw[thick,postaction={decorate},decoration={markings, mark=at position 0.6 with {\arrow{latex}}}](1.8,-0.6)--(1.8,0.6);
\draw[thick,postaction={decorate},decoration={markings, mark=at position 0.7 with {\arrow{latex}}}](1.8,0.6)--(1.8,1.2);
\end{scope}
\end{tikzpicture}
\begin{tikzpicture}[anchor=base,baseline]
\begin{scope}[scale={1}]
\draw[thick,decorate,decoration={complete sines, number of sines=3,amplitude=5pt}](0,0)--(-0.6,0);
\draw[thick,postaction={decorate},decoration={markings, mark=at position 0.6 with {\arrow{latex}}}](0.80,0.60)--(0.0,0.0)node[pos=0.52,above]{};
\draw[thick,postaction={decorate},decoration={markings, mark=at position 0.6 with {\arrow{latex}}}](0,0)--(0.8,-0.6)node[pos=0.52,right]{};
\draw[double,thick,decorate,decoration={complete sines, number of sines=3,amplitude=5pt}](0.82,-0.58)--(1.78,-0.58);
\draw[double,thick,decorate,densely dashed](0.8,0.58)--(1.78,0.58);
\draw[thick,postaction={decorate},decoration={markings, mark=at position 0.6 with {\arrow{latex}}}](0.8,-0.6)--(0.8,0.6);
\draw[thick,postaction={decorate},decoration={markings, mark=at position 0.7 with {\arrow{latex}}}](1.8,-1.2)--(1.8,-0.6);
\draw[thick,postaction={decorate},decoration={markings, mark=at position 0.6 with {\arrow{latex}}}](1.8,-0.6)--(1.8,0.6);
\draw[thick,postaction={decorate},decoration={markings, mark=at position 0.7 with {\arrow{latex}}}](1.8,0.6)--(1.8,1.2);
\end{scope}
\end{tikzpicture}
\begin{tikzpicture}[anchor=base,baseline]
\begin{scope}[scale={1}]
\draw[thick,decorate,decoration={complete sines, number of sines=3,amplitude=5pt}](0,0)--(-0.6,0);
\draw[thick,postaction={decorate},decoration={markings, mark=at position 0.6 with {\arrow{latex}}}](0.0,0.0)--(0.8,0.6)node[pos=0.52,above]{};
\draw[thick,postaction={decorate},decoration={markings, mark=at position 0.6 with {\arrow{latex}}}](0.8,-0.6)--(0,0)node[pos=0.52,left]{};
\draw[double,thick,decorate,decoration={complete sines, number of sines=3,amplitude=5pt}](0.82,0.58)--(1.78,0.58);
\draw[double,thick,decorate,densely dashed](0.8,-0.58)--(1.78,-0.58);
\draw[thick,postaction={decorate},decoration={markings, mark=at position 0.6 with {\arrow{latex}}}](0.8,0.6)--(0.8,-0.6);
\draw[thick,postaction={decorate},decoration={markings, mark=at position 0.7 with {\arrow{latex}}}](1.8,-1.2)--(1.8,-0.6);
\draw[thick,postaction={decorate},decoration={markings, mark=at position 0.6 with {\arrow{latex}}}](1.8,-0.6)--(1.8,0.6);
\draw[thick,postaction={decorate},decoration={markings, mark=at position 0.7 with {\arrow{latex}}}](1.8,0.6)--(1.8,1.2);
\end{scope}
\end{tikzpicture}
\begin{tikzpicture}[anchor=base,baseline]
\begin{scope}[scale={1}]
\draw[thick,decorate,decoration={complete sines, number of sines=3,amplitude=5pt}](0,0)--(-0.6,0);
\draw[thick,postaction={decorate},decoration={markings, mark=at position 0.6 with {\arrow{latex}}}](0.8,0.6)--(0.0,0.0)node[pos=0.52,above]{};
\draw[thick,postaction={decorate},decoration={markings, mark=at position 0.6 with {\arrow{latex}}}](0.0,0.0)--(0.8,-0.6)node[pos=0.52,left]{};
\draw[double,thick,decorate,decoration={complete sines, number of sines=3,amplitude=5pt}](0.82,0.58)--(1.78,0.58);
\draw[double,thick,decorate,densely dashed](0.8,-0.58)--(1.78,-0.58);
\draw[thick,postaction={decorate},decoration={markings, mark=at position 0.6 with {\arrow{latex}}}](0.8,-0.6)--(0.8,0.6);
\draw[thick,postaction={decorate},decoration={markings, mark=at position 0.7 with {\arrow{latex}}}](1.8,-1.2)--(1.8,-0.6);
\draw[thick,postaction={decorate},decoration={markings, mark=at position 0.6 with {\arrow{latex}}}](1.8,-0.6)--(1.8,0.6);
\draw[thick,postaction={decorate},decoration={markings, mark=at position 0.7 with {\arrow{latex}}}](1.8,0.6)--(1.8,1.2);
\end{scope}
\end{tikzpicture}
\end{center}
\vspace{-10pt}
\caption{$\mathcal{O}\left(1/N_{f}\right)$ diagrams of fermionic self-energy $\Sigma_{f}$ and vertex one(two)-loop correction $\Xi_{b}^{(1)} (\Xi_{b}^{(2)})$. Straight arrowed line are the
femionic propagator and wiggly (dashed) line are the emergent gauge boson $\mathcal{B}$ (coulomb $\varphi$) propagators and doubled wiggly (dashed) line are the renormalized propagators.} 
\label{fig6}
\end{figure}

We perform large $N_{f}$ analysis to control the calculation by introducing the flavor number of fermions  $N_{f}$ as a control parameter $1/N_{f}$. The dimension of spinor $\psi$ is extended from $4$ to $4N_{f}$ and vertex coupling constant becomes $g_{b}\rightarrow g_{b}/\sqrt{N_{f}}$. We summarize the conventions for Green's function and self-energies at the one-loop level. \\
\indent First, the bosonic Green's function and self-energy are
\begin{eqnarray}
\mathcal{G_{B}}^{ij}(q_{n},\vec{q}) & = & \left\langle \mathcal{B}^{i}(q)\mathcal{B}^{j}(-q)\right\rangle =\left[\mathcal{G_{B}}_{0}^{ij}(q_{n},\vec{q})^{-1}+\Sigma_{\mathcal{B}}^{ij}(q_{n},\vec{q})\right]^{-1},\\
\mathcal{G}_{\varphi}(q_{n},\vec{q}) & = & \left\langle \varphi(q)\varphi(-q)\right\rangle =\left[\mathcal{G}_{\varphi,0}(q_{n},\vec{q})^{-1}+\Sigma_{\varphi}(q_{n},\vec{q})\right]^{-1}.
\end{eqnarray}
\indent Second, the fermionic Green's function and self-energy to $\mathcal{O}(1/N_{f})$ order are
\begin{eqnarray}
\mathcal{G}_{f}^{\mu\nu}(q_{n},\vec{q}) & = & \left\langle \psi^{\mu}(q)\psi^{\nu}(-q)\right\rangle =\left[\mathcal{G}_{f}^{0,\mu\nu}(q_{n},\vec{q})^{-1}+\Sigma_{f}^{\mu\nu}(q_{n},\vec{q})\right]^{-1},\\
\Sigma_{f}^{\mu\nu}(q_{n},\vec{q}) & = &\frac{1}{N_{f}}\log(\frac{\Lambda}{\mu})\left[\left(-iq_{n}\delta_{\omega}+\frac{\vec{q}^{2}}{2m}\delta_{0}\right)\Gamma^{0}+\sum_{a=1}^{5}d_{a}(\vec{q})\delta_{a}\Gamma^{a}\right],
\end{eqnarray}
where UV(IR) cutoff $\Lambda$($\mu$) are introduced.\\
\indent Third, the vertex functions $\hat{M}_{B^{i},0}=s^{i},\ \hat{M}_{\varphi,0}=I$ are also renormalized with one-loop and two -loop diagrams to $\mathcal{O}(1/N_{f})$ order.
\begin{eqnarray}
\hat{M}_{b} & = & g_{b}\left(\hat{M}_{b,0}+\Xi_{b}^{(1)}+\Xi_{b}^{(2)}\right)=g_{b}\left(1+\frac{1}{N_{f}}\log(\frac{\Lambda}{\mu})\delta_{b}\right)\hat{M}_{b,0},\\
\Xi_{b}^{(1)} & = & \sum_{b'=B,\varphi}\Xi_{b,b'}^{(1)}=\frac{1}{N_{f}}\log(\frac{\Lambda}{\mu})\left(\delta_{b,B}+\delta_{b,\varphi}\right)\hat{M}_{b,0},\\
\Xi_{b}^{(2)} & = & \sum_{b',b"=B,\varphi}\Xi_{b,b',b"}^{(2)}=\frac{1}{N_{f}}\log(\frac{\Lambda}{\mu})\left(\delta_{b,B,B}+\delta_{b,B,\varphi}+\delta_{b,\varphi, B}+\delta_{b,\varphi,\varphi}\right)\hat{M}_{b,0},
\end{eqnarray}
In Fig.\ref{fig6}, we draw all $\mathcal{O}\left(1/N_{f}\right)$ diagrams in our theory. All other diagrams are either higher order than $\mathcal{O}(1/N_{f})$ or zero.
\vspace{-15pt}
\vspace{-10pt}
\subsection{The bosonic self-energies } 
The bosonic self-energies of $\mathcal{B}$ and $\mathcal{\varphi}$ are 
\begin{eqnarray}
\Sigma_{\mathcal{B}}^{ij}(q_{n},\vec{q}) & = & g_{\mathcal{B}}^{2}\int_{\vec{k},k_{n}}\mathrm{Tr}\left(s^{i}G_{f}^{0}(k_{n},\vec{k})s^{j}G_{f}^{0}(k_{n}+q_{n},\vec{k}+\vec{q})\right),\\
\Sigma_{\varphi}(q_{n},\vec{q}) & = & g_{\varphi}^{2}\int_{\vec{k},k_{n}}\mathrm{Tr}\left(G_{f}^{0}(k_{n},\vec{k})G_{f}^{0}(k_{n}+q_{n},\vec{k}+\vec{q})\right).
\end{eqnarray}
We obtain $\delta\Sigma_{B}^{ij}(\vec{q},q_{n})$, $\delta\Sigma_{\varphi}(\vec{q},q_{n})$ by subtracting the UV divergent piece $\Sigma_{\mathcal{B}}^{ij}(0,0)=-\frac{1}{2\pi^{2}}g_{B}^{2}\Lambda\delta^{ij},\ \Sigma_{\varphi}(0,0)=0$.
\begin{eqnarray}
\delta \Sigma_{\mathcal{B}}^{ij}(q_{n},\vec{q})& = & g_{\mathcal{B}}^2\left(a_{\mathcal{B}}^{T}\left|\vec{q}\right|+a_{\mathcal{B}}^{\omega}\sqrt{\left|q_{n}\right|}\right)P_{T}^{ij}(\vec{q}),\\
\delta\Sigma_{\varphi}(q_{n},\vec{q}) & = &g_{\varphi}^2 \left(a_{\varphi}^{q}\left|\vec{q}\right|+a_{\varphi}^{\omega}\sqrt{\left|q_{n}\right|}\right),
\end{eqnarray}
where $a_{\mathcal{B}}^{T}=+0.036,\ a_{\varphi}^{q}=-0.068,\ a_{\mathcal{B}}^{\omega}=+0.040,\ a_{\varphi}^{\omega}=0$ and $P_{T}^{ij}(\vec{q})=(\delta^{ij}-\hat{q}^{i}\hat{q}^{j})$ used. 
\vspace{-10pt}
\subsection{The fermionic self-energies } 
We provide details of calculations of the fermionic self-energy and the vertex correction by gauge and Coulomb fluctuation. Near QCP, we use the bosonic propagators, ${\mathcal{G}_{\mathcal{B},T}}^{ij}(q_{n},\vec{q})=\delta\Sigma_{\mathcal{B},T}^{ij}(q_{n},\vec{q})^{-1}$ and ${\mathcal{G}_{\varphi}}(q_{n},\vec{q})=\delta\Sigma_{\varphi,T}(q_{n},\vec{q})^{-1}$. The fermionic self-energy is
\begin{eqnarray}
\Sigma_{f}(q_{n},\vec{q})&=&-\frac{{g_{B}}^{2}}{N_{f}}\int_{\vec{k},k_{n}} s^{i}G_{f}^{0}(k_{n}+q_{n},\vec{k}+\vec{q}) s^{j}\mathcal{G}_{\mathcal{B}}^{ij}(k_{n},\vec{k})-\frac{{g_{\varphi}}^{2}}{N_{f}}\int_{\vec{k},k_{n}} G_{f}^{0}(k_{n}+q_{n},\vec{k}+\vec{q}) \mathcal{G}_{\varphi}(k_{n},\vec{k}),
\end{eqnarray}
where $\delta_{\omega}=+0.194,\ \delta_{0}=+ 0.360,\ \delta_{a}=-0.083$ for the most symmetric condition. We find that the PHS breaking term $\delta_{0}$ is generated  and even larger than $\delta_{1,2}$, even though we initially impose the PHS. We also consider several sets of bare parameters and check that generation of the PHS breaking term is generic as summarized in Table \ref{table6}. If a bare parameter condition guarantees a nodal point fermionic excitation, our analysis is valid and the PHS becomes broken.
\begin{table}[H]
\begin{centering}
\renewcommand{\arraystretch}{1.2}
\begin{tabular}{C{1cm}|C{3cm} |C{3cm}}
\hline 
\hline
Case & Bare parameters &  Log corrections\tabularnewline
\hline
(i)&\thead{$c_{0}=0, c_{1}=c_{2}=1$\\(in the Main text)}  & \thead{$\delta_{0}=+0.360$\\$\delta_{1,2}=-0.083$} \tabularnewline \hline
(ii)&\thead{$c_{0}=0.1,$\\ $c_{1}=c_{2}=1$}  & \thead{$\delta_{0}=+0.360$\\$\delta_{1,2}=-0.084$} \tabularnewline \hline

(iii)&\thead{$c_{0}=0,$\\ $c_{1}=1.1,c_{2}=0.9$}   & \thead{$\delta_{0}=+0.351$\\$\delta_{1}=-0.066$\\$\delta_{2}=-0.177$} \tabularnewline \hline
\hline
\end{tabular}\caption{The logarithmic correction terms of the fermion self-energies with the various bare parameters in the large $N_f$ calculation. The three different cases are presented: (i) The most symmetric, (ii) PHS is weakly broken, (iii) SO(3) symmetry is weakly broken case. The corrections of $c_{0}$ are generically finite and larger than $c_{1,2}$.  }
\label{table6}
\par\end{centering}
\end{table}

We emphasize that the condition of the spin-ice manifold $(\nabla \cdot \vec{B} =0)$ is essential to break PHS. The absence of the condition forbids the PHS breaking. In other words, conventional ferromagnetic fluctuations as in the previous work \cite{PhysRevB.92.035137} does not induce the PHS breaking. Our calculations demonstrate the interplay of the spin-ice manifold and conduction electrons is the key factor of our Multiscale quantum criticality.

\subsection{The vertex corrections}
\indent Now we consider the vertex corrections of $\mathcal{B}$ and $\varphi$.
\begin{eqnarray}
\Xi_{B^{i}}^{(1)} & = &\left[g_{B}^{2}\int_{q,q_{n}} s^{j}G_{f}^{0}(q_{n},\vec{q}) s^{i}G_{f}^{0}(q_{n},\vec{q})s^{k}\mathcal{G}_{\mathcal{B}}^{j,k}(q_{n},\vec{q}) +g_{\varphi}^{2}\int_{q,q_{n}} G_{f}^{0}(q_{n},\vec{q}) s^{i} G_{f}^{0}(q_{n},\vec{q}) \mathcal{G}_{\varphi}(q_{n},\vec{q})\right](s^{i})^{-1},
 \\
 \Xi_{B^{i}}^{(2)} & =&-\left[\sum_{b',b",\eta=\pm1}g_{b'}^{2} g_{b"}^{2}\int_{\begin{subarray}{l}\vec{q}_{1},{q_{1}}_{n}\\\vec{q}_{2},{q_{2}}_{n}\end{subarray}} \mathrm{Tr} \left[ G_{0,1}^{f} \hat{M}_{b'} G_{0,1+\eta 2}^{f} \hat{M}_{b"} G_{0,1}^{f} s^{i} \right] \mathcal{G}_{b',2}\hat{M}_{b'} G_{0,2}^{f } \hat{M}_{b"} \mathcal{G}_{b",2}\right](s^{i})^{-1},\\
\Xi_{\varphi}^{(1)} & = &\left[g_{B}^{2}\int_{q,q_{n}} s^{j}G_{f}^{0}(q_{n},\vec{q}) G_{f}^{0}(q_{n},\vec{q})s^{k}\mathcal{G}_{\mathcal{B}}^{j,k}(q_{n},\vec{q})+g_{\varphi}^{2}\int_{q,q_{n}} G_{f}^{0}(q_{n},\vec{q}) G_{f}^{0}(q_{n},\vec{q}) \mathcal{G}_{\varphi}(q_{n},\vec{q})\right],\\
\Xi_{\varphi}^{(2)} & =&-\left[\sum_{b',b",\eta=\pm1}g_{b'}^{2} g_{b"}^{2}\int_{\begin{subarray}{l}\vec{q}_{1},{q_{1}}_{n}\\\vec{q}_{2},{q_{2}}_{n}\end{subarray}} \mathrm{Tr} \left[ G_{0,1}^{f} \hat{M}_{b'} G_{0,1+\eta 2}^{f} \hat{M}_{b"} G_{0,1}^{f} \right] \mathcal{G}_{b',2}\hat{M}_{b'} G_{0,2}^{f} \hat{M}_{b"} \mathcal{G}_{b",2}\right],
\end{eqnarray}
where $\delta_{\mathcal{B}}=-0.077$ and $\delta_{\mathcal{\varphi}}=+0.194$ for the most symmetric condition. We summarize all the results of femionic self-energy and vertex correction with and without $\varphi$ in Table \ref{table4}. The scaling dimensions of the coupling constants are
\begin{eqnarray}
\left[ \alpha \right]&=&3-2\Delta_{\psi}+\frac{\delta_{\omega}}{N_{f}},\\
\left[c_{a}\right]&=&z+1-2\Delta_{\psi}+\frac{\delta_{a}}{N_{f}},\ \ \ \ \ \ \ \ \ \ \ \ (\mathrm{for}\ \ a=1\cdots 5)\\
\left[ g_{b} \right]&=&z+3-\Delta_{b}-2\Delta_{\psi}+\frac{\delta_{b}}{N_{f}},\ \ \ \ \ (\mathrm{for}\ \ b=B,\varphi).
\end{eqnarray} 
After substituting the values in Table \ref{table4} and comparing $[g_{B}]$ and the results with insertion of $\vec{\tilde{s}}$, we conclude that $\left[s\right]>\left[\tilde{s}\right]$ and the coupling $\vec{\tilde{s}}$ is more relevant than $\vec{s}$. This is inconsistent with the initial condition of fixed point, thus this fixed point is unstable.
\begin{table}[H]
\begin{centering}
\renewcommand{\arraystretch}{1.2}
\begin{tabular}{C{1.5cm} |C{2cm}| C{2cm}}
\hline 
\hline
Log corr. & W/O $\varphi$ & W/ $\varphi$\tabularnewline
\hline
$\delta_{\omega}$ & +0.194 & +0.194\tabularnewline
$\delta_{0}$ & +0.360 & +0.360\tabularnewline
$\delta_{a}$ & -0.157 & -0.083\tabularnewline\hline
$\delta_{\mathcal{B}}$ & -0.185 & -0.077\tabularnewline
$\delta_{\mathcal{\varphi}}$ & $\cdot$ & $+0.194$\tabularnewline
\hline 
\hline
\end{tabular}\caption{The logarithmic correction of fermion and vertex diagram without (with) long-range Coulomb $\varphi$.}
\label{table4}
\par\end{centering}
\end{table}

\section{Landau-damping term}
\indent Here we calculate the bosonic self-energy by using the femionic Green's function, $G_{f}^{0}(q_{n},\vec{q})^{-1}=-iq_{n}+d_{a}(\vec{q})\Gamma^{a}-\epsilon_{F} $ where the parameter $\epsilon_{F}$ which is associated with size of electron-hole pockets is introduced. For $\sqrt{\left|\Omega_{n}\right|}\ll\left|\vec{q}\right|\ll\sqrt{\left|\epsilon_{F}\right|}$, the bosonic self-energy is 
\begin{eqnarray}
\delta \Sigma^{ij}_{\mathcal{B},T}\left(\vec{q},i\Omega_{n};\epsilon_{F}\right)&= &
\gamma_{T}\frac{\left|\Omega_{n}\right|}{\left|\vec{q}\right|}(\delta_{ij}-\hat{q}^{i}\hat{q}^{j}),
 \end{eqnarray}
where $\gamma_{T}=0.025$. Thus, the critical boson has the dynamics similar to the Hertz-Millis theory. ($z=3$,  $\nu=1/2$)

\
\section{Ferroelectric Phase Transition associated with inversion symmetry}
Here we present the ferroelectric quantum phase transition between fractionalized Luttinger semi-metals(F-LSM) and fractionalized ferroelctric semi-metal(fFE-SM). All the calculations are similar to the ones of the above magnetic transition with minor modifications.
\begin{eqnarray}
\mathcal{G_{E}}^{ij}(k_{n},\vec{k}) & = & \left\langle \mathcal{E}^{i}(k)\mathcal{E}^{j}(-k)\right\rangle =\left[\mathcal{G_{E}}_{0}^{ij}(k_{n},\vec{k})^{-1}+\Sigma_{\mathcal{E}}^{ij}(k_{n},\vec{k})\right]^{-1},
\end{eqnarray}
\vspace{-10pt}
\begin{eqnarray}
\Sigma_{\mathcal{E}}^{ij}(q_{n},\vec{q}) & = &- g_{E}^{2}\int_{\vec{k},k_{n}}\mathrm{Tr}\left(\hat{M}_{E}^{i}(\vec{k},\vec{q})G_{f}^{0}(k_{n},\vec{k})\hat{M}_{E}^{j}(\vec{k}+\vec{q},-\vec{q})G_{f}^{0}(k_{n}+q_{n},\vec{k}+\vec{q})\right).
\label{EQ36}
\end{eqnarray}
We consider the two most relevant coupling matrices with $\mathcal{E}$, $\hat{M}_{E_{1}}(\vec{k},\vec{q})= ig_{E,1}\vec{q}$ and $\hat{M}_{E_{2}}(\vec{k},\vec{q})=g_{E,2}(2\vec{k}+\vec{q})\times \vec{s}$ . Remark that the negative sign in (\ref{EQ36}) is originated from the sign of the Euclidean action, which can be manifested by comparing  (\ref{MA}) and (\ref{EA}).\\
\indent At the one-loop level, we calculate the bosonic-self energy with the no-electric monopole condition, $\nabla \cdot \mathcal{E}=0$. Due to the transverse condition, $\Sigma_{\mathcal{E}_{1},T}^{ij}(q_{n},\vec{q})$ always vanishes. 
With a nonzero ferroelectric coupling $g_{E,2}$, the dynamics of $\mathcal{E}$ field is modified as
 \begin{eqnarray}
 \left(\epsilon_{0}\delta_{ij}\right)\frac{\mathcal{E}^{i}(-q)\mathcal{E}^{j}(q)}{2}&\rightarrow& \left(\epsilon_{0}\delta_{ij}+\Sigma^{ij}_{\mathcal{E}_{2}}(0,0;\Lambda)+\delta
\Sigma^{ij}_{\mathcal{E}_{2}}(q_{n},\vec{q};\Lambda)\right)\frac{\mathcal{E}^{i}(-q)\mathcal{E}^{j}(q)}{2},
\end{eqnarray}
With SO(3) and PHS, the results are
 \begin{eqnarray}
\delta\Sigma^{ij}_{\mathcal{E}_{2}}(q_{n},\vec{q};\Lambda)&=&\Sigma^{ij}_{\mathcal{E}_{2}}(q_{n},\vec{q};\Lambda)-\Sigma^{ij}_{\mathcal{E}_{2}}(0,0;\Lambda),\\
&=&-0.0152g_{E_{2}}^{2}\Lambda^{2}\left|\vec{q}\right|\left(\delta^{ij}-\frac{q^{i}q^{j}}{\vec{q}^2}\right).
\label{EQ39}
\end{eqnarray}
with $\Sigma^{ij}_{\mathcal{E}_{2}}(0,0;\Lambda)=+\frac{2}{3\pi^{2}}g_{E_{2}}^2\Lambda^{3}\delta^{ij}$. 
 The negative sign of (\ref{EQ39}) implies that the gauge fluctuation $\mathcal{E}$ may be unstable at the one-loop-level. \\

\section{Summary of critical exponents and comparison with experiments}
We summarize our critical exponents and compare them with experimentally observed ones. 
The previous work of the magnetic Gruneisen parameters in $\mathrm{Pr_{2}Ir_{2}O_{7}}$ [\onlinecite{Tokiwa2014}] reported that 
\begin{eqnarray}
\Gamma_{H}=\frac{1}{h_{ext}}\mathcal{F}_{\mathrm{exp}}\left(\frac{T^{3/4}}{h_{ext}}\right), \nonumber
\end{eqnarray}
giving $z \nu_{ext}=4/3$. $\nu_{ext}$ is for for the external magnetic field and $z$ is the dynamic critical exponent.
Note that $\nu_{ext}$ corresponds to $2\nu$ of the literature [\onlinecite{Tokiwa3}]. 
Our Multiscale quantum criticality is characterized by the three sets of the critical exponents, 
\begin{eqnarray}
&&z_1 \nu_{ext,1} = 1/2 \quad \quad \quad \quad \quad {\rm emergent \, photons,} \nonumber \\
&&z_2 \nu_{ext,2} = 1 + O(1/N_f) \quad {\rm F-LSM,}  \nonumber \\
&&z_3 \nu_{ext,3} = 3/4 \quad \quad \quad \quad \quad {\rm Hertz-Millis\, QCP}. \nonumber 
\end{eqnarray}
We stress that our theory has a characteristic mechanism to break the particle-hole symmetry from the Kondo lattice coupling between the Luttinger semi-metal and quantum spin-ice, which naturally explains the appearance of Fermi-pockets at low temperatures with multi scaling behaviors though the calculated critical exponents are not exactly  same as the experimentally-determined value.

%

\end{document}